\documentclass[prd,twocolumn,floatfix,altaffilletter,tightenlines,showpacs,showkeys,
               notitlepage,preprintnumbers,nofootinbib,superscriptaddress]{revtex4-1}

\pdfoutput=1
\usepackage[sort&compress]{natbib}

\usepackage{latexsym}
\usepackage{amsthm}
\usepackage{float}
\usepackage{amsmath,amssymb}
\usepackage{graphicx,grffile}
\usepackage{fancybox,feynmp}
\usepackage{indentfirst}
\usepackage{epsfig}
\usepackage{epsfig}
\usepackage{dcolumn}
\usepackage{bm}
\usepackage{slashed}
\usepackage{subcaption}
\usepackage{cancel}
\usepackage{color}
\usepackage{tabularx}
\usepackage[titletoc]{appendix}

\graphicspath{{./Figures/}}

\usepackage{epstopdf}
\usepackage{epstopdf}

\usepackage{booktabs, tabularx}

\usepackage[utf8]{inputenc}
\usepackage{array,booktabs,calc}
\usepackage{multirow}

\newcolumntype{N}{>{\centering\arraybackslash}m{4cm}}
\newcolumntype{G}{>{\bfseries\centering\arraybackslash}m{3cm+6\tabcolsep}}
\newcolumntype{M}[1]{>{\centering\arraybackslash}m{#1}}

\newcommand{\eqnref}[1]{Eq.~(\ref{eqn:#1})}
\newcommand{\eqnsref}[2]{Eqs.~(\ref{eqn:#1}) and (\ref{eqn:#2})}
\newcommand{\secref}[1]{Sec.~\ref{sec:#1}}
\newcommand{\secsref}[2]{Secs.~\ref{sec:#1} and \ref{sec:#2}}
\newcommand{\subsecref}[1]{Subsec.~\ref{subsec:#1}}

\newcommand{\appref}[1]{Appendix~\ref{sec:#1}}
\newcommand{\figref}[1]{Fig.~\ref{fig:#1}}

\newcommand{\tableref}[1]{Table~\ref{table:#1}}

\usepackage[dvipsnames]{xcolor}

\usepackage[colorlinks=true,citecolor=purple,linkcolor=blue]{hyperref}
\usepackage[noabbrev]{cleveref} 

\begin{document}

\preprint{MITP/20-052}

\title{Probing new $U(1)$ gauge symmetries via exotic $Z \to Z' \gamma$ decays}

\author{Lisa Michaels}
\email{lisa.michaels@uni-mainz.de}
\affiliation{PRISMA$^+$ Cluster of Excellence \& Mainz Institute for
  Theoretical Physics, Johannes Gutenberg University, 55099 Mainz,
  Germany}

\author{Felix Yu}
\email{yu001@uni-mainz.de}
\affiliation{PRISMA$^+$ Cluster of Excellence \& Mainz Institute for
  Theoretical Physics, Johannes Gutenberg University, 55099 Mainz,
  Germany}

\date{\today}

\begin{abstract}
New $U(1)$ gauge theories involving Standard Model (SM) fermions typically require additional electroweak fermions for anomaly cancellation.  We study the non-decoupling properties of these new fermions, called anomalons, in the $Z-Z'-\gamma$ vertex function, reviewing the connection between the full model and the effective Wess-Zumino operator.  We calculate the exotic $Z \to Z' \gamma$ decay width in $U(1)_{B-L}$ and $U(1)_B$ models, where $B$ and $L$ denote the SM baryon and lepton number symmetries.  For $U(1)_{B-L}$ gauge symmetry, each generation of SM fermions is anomaly free and the exotic $Z \to Z_{BL}' \gamma$ decay width is entirely induced by intragenerational mass splittings.  In contrast, for $U(1)_B$ gauge symmetry, the existence of two distinct sources of chiral symmetry breaking enables a heavy, anomaly-free set of fermions to have an irreducible contribution to the $Z \to Z_B' \gamma$ decay width.  We show that the current LEP limits on the exotic $Z \to Z'_B \gamma$ decay are weaker than previously estimated, and low-mass $Z'_B$ dijet resonance searches are currently more constraining.  We present a summary of the current collider bounds on $U(1)_B$ and a projection for a TeraZ factory on the $Z \to Z_B' \gamma$ exotic decay, and emphasize how the $Z \to Z' \gamma$ decay is emblematic of new anomalous $U(1)$ gauge symmetries.

\end{abstract}


\maketitle

\section{Introduction}
\label{sec:Introduction}

The Standard Model (SM) exhibits a chiral electroweak gauge symmetry under which bare mass terms for the elementary quarks and charged leptons are forbidden.  Correspondingly, gauge anomaly cancellation imposes conditions on the gauge quantum numbers of the fermions, without which the SM would suffer a failure of renormalizability~\cite{Preskill:1990fr}.  Charging the SM fermions under an additional gauge symmetry, such as baryon number or lepton number~\cite{FileviezPerez:2010gw}, imposes more anomaly cancellation conditions which necessitates introducing new fermion fields with electroweak quantum numbers. A prime advantage of such a construction is the prediction of new $Z'$ gauge bosons, which give promising dijet and dilepton resonances at colliders~\cite{Carena:2004xs, Langacker:2008yv, Dobrescu:2013cmh}.  New $U(1)$ symmetries are ubiquitous in beyond the Standard Model (BSM) physics, since they are motivated by numerous extensions of the SM, such as grand unified models that gauge the SM baryon minus lepton number~\cite{Langacker:2008yv}, the individual baryon number or lepton number global symmetries~\cite{Carena:2004xs, Anastasopoulos:2006cz, FileviezPerez:2010gw, FileviezPerez:2011pt, Dobrescu:2013cmh} or providing portals to dark sectors, as in Refs.~\cite{Cui:2017juz, Dror:2017ehi, Arcadi:2017jqd, Dror:2017nsg, Ismail:2017ulg, Ismail:2017fgq, Dror:2018wfl}.

The focus of this work is exploring the phenomenology of new $U(1)$ gauge extensions of the SM, particularly when the $Z'$ boson is light as well as when the $U(1)$ symmetry is an anomalous global symmetry of the SM.  Recent work on mapping out the possible $U(1)$ gauge extensions to the SM include Refs.~\cite{Batra:2005rh, Allanach:2018vjg, Costa:2019zzy, Costa:2020dph}.  When gauging baryon number, the anomaly coefficients $SU(2)_L^2 \times U(1)_B$ and $U(1)_Y^2 \times U(1)_B$ are nonzero after summing over the SM quark fields, requiring the introduction of new electroweak fields, called anomalons, for ultraviolet (UV) consistency.  Correspondingly, the anomalons must be massless in the unbroken phase when all spontaneous breaking of all gauge symmetries is turned off.  Moreover, as a consequence of their chiral couplings, the virtual effects of the anomalons in loop processes is entirely reminiscent of the behavior of the SM fermions in loop-induced Higgs processes.  For instance, the heavy mass limit for loops of anomalon fields in the $Z-Z'-\gamma$ vertex does not decouple but instead approaches a nonzero constant.  This non-decoupling behavior is familiar from the established exclusion of a pure four generation SM, as the heavy fourth generation quarks would dramatically enhance the overall cross section of the 125 GeV Higgs boson by a factor of nine~\cite{Kribs:2007nz, Khachatryan:2016vau}.  

In the present context, where we consider the interaction of three distinct neutral current vectors mediated at 1-loop by fermions with vector and axial-vector couplings, the non-decoupling limit of the virtual fermions gives a contribution to the effective Wess-Zumino term~\cite{Wess:1971yu}.  Specifically, we calculate the $Z-Z^\prime-\gamma$ interaction vertex induced by fermions, where the $Z$ and $\gamma$ are the usual SM gauge bosons and the $Z^\prime$ boson corresponds to a new, spontaneously broken $U(1)$ gauge symmetry.  Using a UV complete model, we elucidate the matching condition from heavy anomalons and the effective Wess-Zumino operator, particularly since both, the $Z$ and the $Z^\prime$ symmetries are spontaneously broken.

Knowing the $Z-Z^{\prime}-\gamma$ vertex is the critical ingredient to calculating the exotic $Z \to Z^{\prime}\gamma$ decay, which is a new and attractive channel for studying light $Z^{\prime}$ bosons.  The corresponding vertex diagrams are reminiscent of the standard Adler-Bell-Jackiw anomaly calculation~\cite{Adler:1969gk, Bell:1969ts}, but in order to calculate the decay width, we require the full vertex structure and not only the divergences of each current~\cite{Dedes:2012me}.

After performing the vertex calculation, we discuss the phenomenology of the anomalons in the UV-complete theory and the effective theory in matching to a Wess-Zumino term.  Additionally, as the anomalon fields exhibit non-decoupling in loop-induced scalar decays, we can also constrain the parameter space of realistic gauged $U(1)$ models by measurements of the observed Higgs boson.

In~\secref{models}, we review the two concrete models of $U(1)_{B-L}$ and $U(1)_B$ gauge symmetries and their respective anomaly cancellation conditions.  In~\secref{ZZpgamma}, we calculate the general $Z-Z'-\gamma$ vertex function by imposing the appropriate Ward-Takahashi identities (WIs) on all external currents.  In~\secref{direct}, we discuss the current collider constraints on electroweak charged anomalons, applicable to general $U(1)$ extensions of the SM where the anomalons form vector-like representations under the SM gauge groups.  In~\secref{couplingmass}, we also present the current suite of constraints in the $g_X$ vs.~$m_{Z'_B}$ plane for the $U(1)_B$ model.  We conclude in~\secref{conclusions}.  We give a comparison to an effective operator approach and a critique of the Goldstone boson equivalence implementation in the $Z \to Z'_B \gamma$ decay calculation in~\appref{comparison}.

\section{Models of additional $U(1)$ gauge symmetries}
\label{sec:models}

Two particularly interesting models of additional $U(1)$ gauge symmetries are the familiar $U(1)_{B-L}$ and the anomalous $U(1)_B$ symmetries, where $B$ is baryon number and $L$ is lepton number.  All SM quarks carry $B$ charge of $1/3$ and SM leptons carry $L$ charge of $1$.  The gauged $B-L$ symmetry has been studied extensively in the literature to prevent proton decay in grand unified theories~\cite{Langacker:2008yv}; moreover, with the introduction of right-handed neutrinos for neutrino masses, gauged $B-L$ is also anomaly-free.

From a top-down view, the global symmetries of the SM with Dirac neutrino masses is $U(1)_B \times U(1)_L$: we can thus gauge an arbitrary subgroup of this product symmetry group without affecting the SM Yukawa structure.  All distinct possibilities can be parametrized by $U(1)_L$ and $U(1)_{B - xL}$, where $x$ is a real multiplicative factor~\cite{Carena:2004xs}.  For any choice of $x \neq 1$, the total contribution of SM fields to $SU(2)_L^2 \times U(1)_{B - xL}$ and $U(1)_Y^2 \times U(1)_{B - xL}$ anomalies is nonzero.  For example, for $x = 0$, we have $SU(2)_L^2 \times U(1)_B = 3/2$ and $U(1)_Y^2 \times U(1)_B = -3/2$~\cite{FileviezPerez:2010gw}.  Thus, we must add new electroweak matter to cancel the mixed anomalies and also ensure all other gauge anomalies vanish.  As mentioned in the introduction, since the anomalon fields do not obey decoupling, they will contribute to effective operators at low energies, which can be probed in $Z$ and Higgs decays.

We emphasize that the essential distinction between the first category of anomaly-free $U(1)$ gauge symmetries (such as $U(1)_{B-L}$) and the second category of anomalous symmetries (such as $U(1)_B$) is whether there are one or two scales of chiral symmetry breaking.  In the first category, the $Z'$ boson mass is clearly independent of the SM masses: a simple Abelian Higgs model can serve as the UV completion of the massive $Z'$ boson, where the corresponding Higgs boson can be made heavy by a large quartic coupling and not appear in the mass spectrum.

In the second category, the $U(1)$-breaking vacuum expectation value (vev) of the underlying $U(1)$-charged Higgs field plays a critical role by giving mass to both the $Z'$ boson and the anomalon fields. In effect, hierarchies between the Yukawa couplings determining the physical fermion masses and the gauge coupling determining the $Z'$ boson mass dictate the resulting separation between the particle species.  Since these couplings are renormalizeable, this mass splitting is stable under renormalization group evolution, giving rise to the possibility that in a new sector of physics, the first kinematically accessible state will be a $Z'$ boson, and the fermionic degrees of freedom are further in the UV.  Nevertheless, in an effective field theory (EFT) description where the anomalons are integrated out~\cite{Preskill:1990fr}, the only possible scale suppression of the higher dimension operators is simply the $U(1)$-breaking vev~\cite{DHoker:1984mif, DHoker:1984izu}.  From this perspective, the non-decoupling behavior of the chiral anomalons is responsible both for the Wess-Zumino term that arises in loop-induced vertex functions of vectors~\cite{Wess:1971yu}, and the low-energy theorem in Higgs physics~\cite{Shifman:1979eb}.


\subsection{Anomaly-free models: Gauged $U(1)_ {\text{B-L}}$ symmetry}
\label{sec:freemodels}

For the classic $B-L$ gauge symmetry, the SM field content is augmented by three electroweak singlet right-handed neutrinos, which are required to cancel the $\sum U(1)_L$ and $U(1)_L^3$ anomalies.  We remark that each generation of fermions satisfies the $U(1)_{B-L}$ anomaly cancellation conditions independently.  As we will see in~\secref{ZZpgamma}, this is why the net contribution of an entire generation of mass degenerate SM fermions vanishes in anomaly-induced processes.  Hence, similarly to the GIM mechanism~\cite{Glashow:1970gm}, the net effect of an anomaly-free set of fermions in chiral anomaly-probing interactions is proportional to a charge-weighted mass difference of the respective fermions.

\subsection{Anomalous Models: Gauged $U(1)_B$ Symmetry}
\label{sec:modelU1B}

Our exemplary model for gauging anomalous global symmetries of the SM is gauged baryon number $U(1)_B$.  Again, all SM quarks carry charge $1/3$, but this gives a nonzero anomaly coefficient to the mixed electroweak anomalies: $\mathcal{A} (SU(2)_L^2 \times U(1)_B) = 3/2$ and $\mathcal{A} (U(1)_Y^2 \times U(1)_B) = -3/2$~\cite{FileviezPerez:2010gw}.  Any set of fermions introduced to cancel these anomalies must necessarily keep the other gauge anomalies zero. Since the mixed anomaly $SU(3)_C^2 \times U(1)_B$ is already zero, the new fermions do not need to carry color charge, but they must carry electroweak charges to cancel the mixed electroweak-baryon number anomalies.

Following Ref.~\cite{Dobrescu:2014fca}, we will hence consider a minimal set of colorless anomalons, denoted by $L_L$, $L_R$, $E_L$, $E_R$, $N_L$, and $N_R$, which mimic the SM leptons in their electroweak quantum numbers.  The gauge charges for the new fermions and the $U(1)_B$ scalar Higgs field $\Phi$ are shown in~\tableref{quantumnumbers}.  Since the new fermions come in vector-like pairs under the $SU(2)_L \times U(1)_Y$ gauge symmetry, there are no new pure electroweak gauge anomalies.  The $L_L$, $L_R$, $E_L$, and $E_R$ fields cancel the $SU(2)_L^2 \times U(1)_B$ and $U(1)_Y^2 \times U(1)_B$ mixed anomalies from the SM quarks but introduce $\sum U(1)_B$ and $U(1)_B^3$ anomalies, which are correspondingly cancelled by $N_L$ and $N_R$.  Note that $N_L$ and $N_R$ can alternatively carry $U(1)_B$ charges $1$ and $-2$, respectively.  These charges also satisfy the trace condition, Tr$(q_B Y) = 0$, summing over all fermions, which is necessary to avoid large $B-Z'_B$ mixing~\cite{Dobrescu:2014fca}, where $B$ is the hypercharge gauge boson.  We write the interactions of the SM quark fields with the $Z'_B$ vector as
\begin{align}
    \mathcal{L} \supset
    g_X \frac{1}{3} Z'_{B\mu} \left( \overline{q} \gamma^\mu q \right) \ ,
\label{eqn:U1BLag}
\end{align}
where $g_X$ is the $U(1)_B$ gauge coupling.  The SM quarks always have vector couplings to the $Z'_B$ boson.

\vspace{0.5cm} 
\begin{table}[tbh!]
\begin{tabular}{c | c | c | c }
Field & $SU(2)_L$ & $U(1)_Y$ & $U(1)_B$ \\ \hline
$L_L$ & 2 & $-1/2$ & $-1$ \\ \hline
$L_R$ & 2 & $-1/2$ & $2$\\ \hline 
$E_L$ & 1 & $-1$ & $2$ \\ \hline
$E_R$ & 1 & $-1$ & $-1$ \\ \hline
$N_L$ & 1 & $0$ & $2$ \\ \hline
$N_R$ & 1 & $0$ &  $-1$ \\ \hline
$\Phi$ & 1 & $0$ & $3$
\end{tabular}
\caption{Quantum numbers of colorless anomalons and the $U(1)_B$ Higgs field $\Phi$, from Ref.~\cite{Dobrescu:2014fca}.}
\label{table:quantumnumbers}
\end{table}

The anomalon masses are protected by the SM electroweak chiral symmetry and the $U(1)_B$ chiral symmetry.  Hence, at least one source of chiral symmetry breaking, either the Higgs vev or the vev of the $U(1)_B$-breaking field $\Phi$, is needed in order to give mass to all of the fermions.

For the model content in~\tableref{quantumnumbers}, we write the Lagrangian as
\begin{align}
\mathcal{L} =& \ \mathcal{L}_{\text{kin}} + \mathcal{L}_{\text{Yuk}} + \mathcal{L}_{\text{scalar}} \,,\\
\mathcal{L}_{\text{kin}} =& \ \bar{L}_L i \slashed{D} L_L + \bar{L}_R i \slashed{D} L_R + \bar{E}_L i \slashed{D} E_L\nonumber \\&+ \bar{E}_R i \slashed{D} E_R + \bar{N}_L i \slashed{D} N_L + \bar{N}_R i \slashed{D} N_R\,, \\
\mathcal{L}_{\text{Yuk}} =& -y_L \bar{L}_L \Phi^* L_R - y_E  \bar{E}_L \Phi E_R - y_N \bar{N}_L \Phi N_R\nonumber  \\  
&- y_1 \bar{L}_L H E_R - y_2 \bar{L}_R H E_L\nonumber \\ &- y_3 \bar{L}_L \tilde{H} N_R - y_4 \bar{L}_R \tilde{H} N_L + \text{ h.c.} \label{eqn:Lyuk} \,, \\
\mathcal{L}_{\text{scalar}} =& \ |D_\mu \Phi|^2 -\mu_{\Phi}^2 |\Phi|^2 - \lambda_{\Phi} |\Phi|^4 - \lambda_{H\Phi} |H|^2 |\Phi|^2 \ .
\end{align}
We will assume $\mu_{\Phi}^2 < 0$, which will trigger spontaneous breaking of $U(1)_B$ at a scale $v_\Phi$, where $\Phi = (v_\Phi + \phi) / \sqrt{2}$. We will also assume the Higgs portal scalar coupling $\lambda_{H\Phi}$ is negligible throughout this work.  Similarly, we will ignore a possible kinetic mixing term between the hypercharge field strength and the baryon-number field strength in the calculations.

The Yukawa Lagrangian in~\eqnref{Lyuk} exhibits two sources for generating masses for the anomalons.  The $y_L$, $y_E$, and $y_N$ couplings become SM vector-like masses for the anomalons once $\Phi$ acquires a vev, since the anomalons come in vector-like pairs under the electroweak gauge symmetry.  The $y_1$, $y_2$, $y_3$ and $y_4$ couplings correspond to SM-like Yukawa terms and mimic the lepton Yukawas of the SM (as well as the role of the SM leptons in cancellation of $B-L$ gauge anomalies).

It is straightforward to write the anomalon contributions to the SM $Z$- and $\gamma$-mediated currents as well as the $U(1)_B$ current.  Adopting the electromagnetic charge $Q = T_3 + Y$, and $L_L = (\nu_L, e_L)^T$, $L_R = (\nu_R, e_R)^T$, the current interactions of the anomalons are 
\begin{widetext}
\begin{align}
\mathcal{L}_{int} =& \ e_\text{EM} A_{\mu}J^{\mu}_{\text{EM}} + \frac{g}{c_W } Z_{\mu} J^{\mu}_Z + g_X Z_{B\mu}^{\prime} J^{\mu}_{Z_{B}^{\prime}} \, , \quad \text{with}\\
J^{\mu}_{\text{EM}} =& \  \bar{e}_L (-1) \gamma^\mu e_L + 
\bar{e}_R (-1) \gamma^\mu e_R + \bar{E}_L (-1) \gamma^\mu E_L + \bar{E}_R (-1) \gamma^\mu E_R \, , \\
J^{\mu}_Z =& \ 
\bar{e}_L \left( \frac{-1}{2} + s_W^2 \right) \gamma^\mu e_L +
\bar{e}_R \left( \frac{-1}{2} + s_W^2 \right) \gamma^\mu e_R +
\bar{E}_L \left( s_W^2 \right) \gamma^\mu E_L +
\bar{E}_R \left( s_W^2 \right) \gamma^\mu E_R 
+ \bar{\nu}_L \frac{1}{2} \gamma^\mu \nu_L + \bar{\nu}_R \frac{1}{2} \gamma^\mu \nu_R \ , \\
J^{\mu}_{Z_B^{\prime}} =& \ 
\bar{e}_L (-1) \gamma^\mu e_L + \bar{e}_R (2) \gamma^\mu e_R +
\bar{E}_L (2) \gamma^\mu E_L + \bar{E}_R (-1) \gamma^\mu E_R 
+ \bar{\nu}_L (-1) \gamma^\mu \nu_L + \bar{\nu}_R (2) \gamma^\mu \nu_R \nonumber \\
+&
\bar{N}_L (2) \gamma^\mu N_L + \bar{N}_R (-1) \gamma^\mu N_R \ ,
\end{align}
\end{widetext}
where $c_W$ and $s_W$ are the cosine and sine of the weak angle.  As alluded to above, these Weyl fermions can be paired into Dirac fermions in two limiting cases: (1) $y_L$, $y_E$, $y_N$ nonzero and the other Yukawas zero, or (2) $y_1$, $y_2$, $y_3$, $y_4$ nonzero and the others zero.  

We note that if the dominant source of the fermion masses follows the second case, then the fermions will greatly impact the Higgs diphoton decay rate and be phenomenologically excluded from the observation of the SM-like nature of the 125~GeV Higgs boson~\cite{Khachatryan:2016vau}.  In particular, the charged matter fields will behave as non-decoupling contributions to the $h \to \gamma \gamma$ decay.  On the other hand, if the dominant source of the fermion masses arises from the $y_L$, $y_E$, and $y_N$ couplings, the effects on Higgs observables are diluted and can be consistent with current measurements of Higgs couplings.  We discuss the constraints from Higgs physics and direct searches of the electroweak anomalons in~\secref{direct}.  

Correspondingly, we will mainly adopt the first case, as suggested by our naming convention, where the fermion mass eigenstates are vector-like under the SM gauge symmetry and the axial-vector couplings to the $Z$ boson vanish.  Explicitly, we have 
\begin{align}
\mathcal{L}_\text{mass} =& 
- m_L \left( 1 + \frac{\phi}{v_\Phi} \right) (\bar{\nu} \nu + \bar{e} e) \\  \nonumber
 &- m_E \left( 1 + \frac{\phi}{v_\Phi} \right) \bar{E} E
- m_N \left( 1 + \frac{\phi}{v_\Phi} \right) \bar{N} N \ ,
\end{align}
where $m_L = y_L v_\Phi / \sqrt{2}$, $m_E = y_E v_\Phi / \sqrt{2}$, and $m_N = y_N v_\Phi / \sqrt{2}$, and $\nu$ and $e$ should not be confused with the SM neutrino and electron.  For simplicity, our numerical analysis in~\secsref{ZZpgamma}{direct} will consider degenerate charged anomalons with mass $M = m_L = m_E$. The neutral gauge currents for these fermion mass eigenstates become
\begin{widetext}
\begin{align}
J^{\mu}_{\text{EM}} =& \  \bar{e} (-1) \gamma^\mu e + \bar{E} (-1) \gamma^\mu E \ , \\
J^{\mu}_Z =&  \ 
\bar{e} \left( \frac{-1}{2} + s_W^2 \right) \gamma^{\mu} e + 
\bar{E} s_W^2 \gamma^{\mu} E +
\bar{\nu} \left( \frac{1}{2} \right) \gamma^{\mu} \nu \ , \\
J^{\mu}_{Z_B^{\prime}} =& \ 
\bar{e} \left( \frac{1}{2} \gamma^\mu + \frac{3}{2} \gamma^\mu \gamma^5 \right) e +
\bar{E} \left( \frac{1}{2} \gamma^\mu - \frac{3}{2} \gamma^\mu \gamma^5 \right) E + 
\bar{\nu} \left( \frac{1}{2} \gamma^\mu + \frac{3}{2} \gamma^\mu \gamma^5 \right) \nu +
\bar{N} \left( \frac{1}{2} \gamma^\mu - \frac{3}{2} \gamma^\mu \gamma^5 \right) N
\ .
\end{align}
\end{widetext}

We emphasize that the vanishing of the axial-vector coupling to the $Z$ boson in this limiting case is intimately tied to the vanishing of the Yukawa couplings $y_1$ to $y_4$, and as such, these mass eigenstate anomalons are chiral under $U(1)_B$ and vector-like under the EW symmetry.  In general, a nonzero axial-vector coupling between a fermion $f$ and a neutral gauge boson $V$ leads to perturbative unitarity violation in $f \bar{f} \to VV$ scattering, which is cured by the $s$-channel Higgs insertion~\cite{Appelquist:1987cf}.  Aside from the role of the axial-vector coupling in perturbative unitarity violation, we will focus on the role of the axial-vector coupling in determining the appropriate WIs in triangle diagram calculations of three gauge boson vertices.

\section{The $Z \to Z' \gamma$ vertex}
\label{sec:ZZpgamma}

\subsection{The generic vertex structure}
\label{subsec:vertex}

We now calculate the partial width for an exotic decay of the SM $Z$ boson decaying to a $Z'$ boson and a photon, where the loop is mediated by fermions. Note that for the partial width mediated by one fermion, the anomaly is certainly not cancelled and the result has to depend on the anomaly prescription.  We consider the general situation where the intermediate fermions have vector and axial-vector couplings to each massive gauge boson, $Z$ and $Z'$.  Since the photon mediates an unbroken, non-chiral $U(1)$ gauge symmetry, its coupling is necessarily vector-like.  We show the two diagrams in~\figref{triangle-diagrams}.

The corresponding matrix elements are given by\footnote{We remark that, in the mass basis, the fermions may have flavor-changing neutral current couplings if their masses arise from both chiral sources of electroweak and $U(1)_B$ breaking. The corresponding matrix elements would then involve mixing angles and triangle diagrams with two different fermions as intermediate states.}
\begin{widetext}
\begin{figure*}[th]
 \begin{center}
  \includegraphics[width=\textwidth]{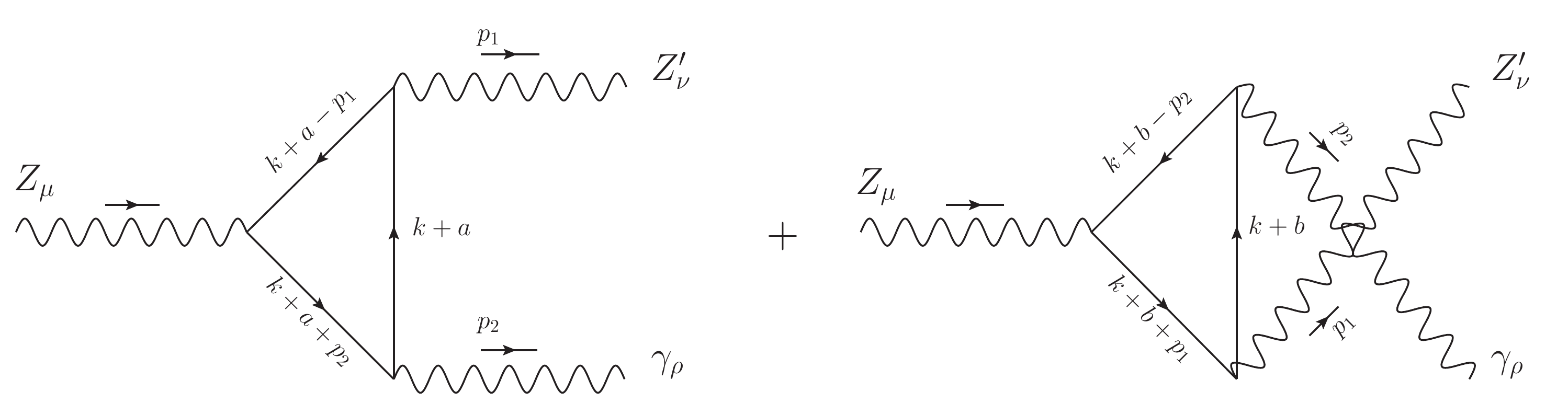}
 \end{center}
 \caption{The triangle diagrams corresponding to the $Z-Z'-\gamma$ vertex function.}
\label{fig:triangle-diagrams}
\end{figure*}

\begin{align}
\label{eqn:ME1}
i \mathcal{M}_1 &= \dfrac{-Qe_\text{EM} g g_X}{(2 \pi)^4} \varepsilon_\mu (p_1 + p_2) \varepsilon_\nu^* (p_1) \varepsilon_\rho^* (p_2) \times \nonumber \\
&\int d^4 k \text{ Tr} \left[ 
(g_{v}^Z \gamma^\mu + g_{a}^Z \gamma^\mu \gamma^5) \dfrac{
    \slashed{k} + \slashed{a}- \slashed{p}_1  + m}{(k+a-p_1)^2 - m^2} (g_{v}^{Z'} \gamma^\nu + g_{a}^{Z'}
  \gamma^\nu \gamma^5) \dfrac{ \slashed{k} + \slashed{a} + m}{(k +
    a )^2 - m^2} \gamma^\rho \dfrac{\slashed{k} + \slashed{a} + \slashed{p}_2 +
     m}{(k + a + p_2)^2 - m^2} \right] \\
i \mathcal{M}_2 &= \dfrac{-Qe_\text{EM} g g_X}{(2 \pi)^4} \varepsilon_\mu (p_1 + p_2) \varepsilon_\nu^* (p_1) \varepsilon_\rho^* (p_2) \times \nonumber \\
\label{eqn:ME2}
&\int d^4 k \text{ Tr} \left[ 
(g_{v}^Z \gamma^\mu + g_{a}^Z \gamma^\mu \gamma^5) \dfrac{
    \slashed{k} + \slashed{b} - \slashed{p}_2+ m}{(k + b-p_2)^2 - m^2} \gamma^\rho \dfrac{ \slashed{k} + \slashed{b} + m}{(k + b)^2 - m^2} (g_{v}^{Z'} \gamma^\nu +
  g_{a}^{Z'} \gamma^\nu \gamma^5) \dfrac{ \slashed{k} + \slashed{b} + \slashed{p}_1 + m}{(k + b + p_1 )^2 - m^2}
  \right]
\end{align}
\end{widetext}
where $Q$ and $m$ are the electric charge and the mass of the fermion in the loop, $g_v$, $g_a$ are the vector and axial coupling factors to the denoted massive $Z$ and $Z'$ gauge bosons, $p_1$ and $p_2$ are the external outgoing momenta and $k$ is the loop momentum, as depicted in Fig.~\ref{fig:triangle-diagrams}. We introduce the arbitrary constant four-vectors $a$ and $b$ as possible shifts in each diagram because the finite result from the cancellation of the divergent integrals depends on the choice of these possible shifts~\cite{Weinberg:1996kr}.  The explicit choices of $a$ and $b$ are generally fixed by applying external physical conditions, as first applied in Ref.~\cite{Rosenberg:1962pp}. From Lorentz and parity symmetry, the most general expression for the vertex function is the sum of the following form factors~\cite{Dedes:2012me},
\begin{widetext}
\begin{align}
\label{eqn:Genvtx}
\Gamma^{\mu \nu \rho} &(p_1, p_2; w, z) = \\
& F_1 (p_1, p_2) \epsilon^{\nu\rho |p_1| |p_2|} p_1^{\mu}\,  + 
F_2 (p_1, p_2) \epsilon^{\nu\rho |p_1| |p_2|} p_2^{\mu}\,  +
F_3 (p_1, p_2) \epsilon^{\mu\rho |p_1| |p_2|} p_1^{\nu}\,  +
F_4 (p_1, p_2) \epsilon^{\mu\rho |p_1| |p_2|} p_2^{\nu}\,   \nonumber \\ +
&F_5 (p_1, p_2) \epsilon^{\mu\nu |p_1| |p_2|} p_1^{\rho}\,  +
F_6 (p_1, p_2) \epsilon^{\mu\nu |p_1| |p_2|} p_2^{\rho}\, +
G_1 (p_1, p_2; w) \epsilon^{\mu\nu\rho\sigma} p_{1\sigma} + 
G_2 (p_1, p_2; z) \epsilon^{\mu\nu\rho\sigma} p_{2\sigma}\nonumber \ ,
\end{align}
\end{widetext}
where $\epsilon^{\nu\rho |p_1| |p_2|} = \epsilon^{\nu \rho \alpha \beta} p_{1\alpha} p_{2\beta}$, etc., we set $b = -a$ to avoid a non-chiral anomaly~\cite{Weinberg:1996kr}, and $a$ has been reexpressed in terms of the external momenta, $a^\mu=z\, p_1^\mu + w\, p_2^\mu$ with constant scalar prefactors $w$ and $z$.  The six form factors $F_1$ to $F_6$ are all finite and hence can be calculated in any regularization prescription unambiguously: they are $w$- and $z$-independent.  Moreover, because of the linear dependence of vectors in a four-dimensional space, two of these can be eliminated by using the identity~\cite{Dedes:2012me},
\begin{align}
-p_1^\mu \epsilon^{\nu \rho |p_1| |p_2|}  &= -p_1^{\nu}\, \epsilon^{\mu \rho |p_1| |p_2|}  + p_1^{\rho}\, \epsilon^{\mu \nu |p_1| |p_2|}  \nonumber \\
&+ \epsilon^{\mu \nu \rho \alpha} \left( (p_1 \cdot p_2)\, p_{1\alpha} - p_1^2 \,p_{2 \alpha} \right) \ , \\
-p_2^\mu \epsilon^{\nu \rho |p_1| |p_2|}  &= -p_2^{\nu}\, \epsilon^{\mu \rho |p_1| |p_2|}  + p_2^{\rho}\, \epsilon^{\mu \nu |p_1| |p_2|} \nonumber \\ &- \epsilon^{\mu \nu \rho \alpha} \left( (p_1 \cdot p_2)\, p_{2\alpha} - p_2^2 \,p_{1\alpha} \right) \ ,
\end{align}
which absorb $F_1$ and $F_2$ into redefinitions of the other form factors: we denote the redefined form factors as $F_i'$, $i = 3$ to $6$, and $G_1'$, $G_2'$.  In contrast, the two form factors $G_1'$ and $G_2'$ arise from the cancellation of divergent integrals, and their values depend on the choice of $w$ and $z$.  

From~\eqnref{Genvtx}, the WIs are given by
\begin{align}
   & ( p_{1 \mu} + p_{2 \mu} ) \Gamma^{\mu \nu \rho} =  (G_2'-G_1')\,\epsilon^{\nu \rho |p_1| |p_2|} \,,
    \\
    & -p_{1 \nu} \Gamma^{\mu \nu \rho} = (-F_3'\,p_1^2 - F_4'\,p_1\cdot p_2 +G_2')\epsilon^{\mu \rho |p_1| |p_2|} \,,
    \\
    & -p_{2 \rho} \Gamma^{\mu \nu \rho} = (-F_5'\, p_1\cdot p_2 -F_6'\,p_2^2 +G_1') \epsilon^{\mu \nu |p_1| |p_2|} \,.
\end{align}

Following Ref.~\cite{Dedes:2012me}, which implements the calculation procedure in Refs.~\cite{Adler:1969gk, Rosenberg:1962pp, Treiman:1986ep}, we construct the ambiguous parts of $G_1'$ and $G_2'$ by isolating the divergent piece of the general three-vector vertex associated with the axial vector anomaly.  This divergent piece can be evaluated using a momentum-shift integral identity~\cite{Treiman:1986ep}, which makes the $w$ and $z$-dependent momentum shifts manifest in the definitions of $G_1'$ and $G_2'$.

Moving to the specific case in~\eqnsref{ME1}{ME2}, we calculate the finite form factors of the vertex function in Mathematica~\cite{Mathematica} using Package-X~\cite{Patel:2015tea, Patel:2016fam}.  The WIs become
\begin{widetext}
\begin{align}
    \left( p_{1 \mu} + p_{2 \mu} \right) \Gamma^{\mu \nu \rho} &= \frac{Qe_{\text{EM}} g g_X}{4 \pi^2 c_W} \epsilon^{\nu \rho |p_1| |p_2|} ( (w-z) (g_v^{Z'} g_a^Z + g_v^Z g_a^{Z'}) + 4m^2 g_v^{Z'} g_a^Z C_0(m))\,,
    \label{eqn:wandzWIs1}
    \\
    -p_{1 \nu} \Gamma^{\mu \nu \rho} &= 
    \frac{Qe_{\text{EM}} g g_X}{4 \pi^2 c_W} \epsilon^{\mu \rho |p_1| |p_2|} ( (w-1) (g_v^{Z'} g_a^Z + g_v^Z g_a^{Z'}) - 4 m^2 g_v^Z g_a^{Z'} C_0(m))\,,
    \label{eqn:wandzWIs2}
    \\
    -p_{2 \rho} \Gamma^{\mu \nu \rho} &= 
    \frac{Qe_{\text{EM}} g g_X}{4 \pi^2 c_W} \epsilon^{\mu \nu |p_1| |p_2|} (z+1) (g_v^{Z'} g_a^Z + g_v^Z g_a^{Z'})\, ,
    \label{eqn:wandzWIs3}
\end{align}
\end{widetext}
where
\begin{align}
    C_0(m) \equiv C_0 (0, m_Z^2, m_Z'^2, m, m, m)
\end{align}
is the usual Passarino-Veltman scalar loop function for the triangle loop, following the Package-X convention~\cite{Passarino:1978jh, Patel:2015tea, Patel:2016fam}.

Clearly, each of the WIs in Eqs.~\ref{eqn:wandzWIs1},~\ref{eqn:wandzWIs2}, and~\ref{eqn:wandzWIs3} contain a constant, fermion mass-independent anomaly piece. Moreover, the WIs for the massive gauge bosons also have fermion mass-dependent contributions, but only when the fermion has the corresponding axial-vector coupling.  Since we calculate in the flavor conserving limit, a given mass eigenstate fermion can only have one non-zero axial-vector coupling.

At this point, we could naively adopt the method by Rosenberg~\cite{Rosenberg:1962pp} to set $w$ and $z$ for each fermion such that the vector WIs are vanishing and the anomaly contributes only the axial-vector divergence.  This would be wrong, however, because all fermions in the loop must use the same consistent choice of $w$ and $z$.  The vertex function we study is the first physical case where this mistake would become apparent, because the mixed electroweak-$U(1)_B$ anomaly is cancelled by two distinct chiral sectors of fermions.  On the other hand, for the $B-L$ case, the SM fermions have axial-vector couplings only on the $\mu$ vertex, and thus choosing $w$ and $z$ to make the WIs on the $\nu$- and $\rho$-vertices vanish is consistent for all fermions.

Instead, with a UV-complete model, the WI on each vertex is independent of the choice of $w$ and $z$.  Moreover, when all fermions are massless or otherwise degenerate, the WI on each vertex is also vanishing.  In fact, we can provide an equivalent condition for an anomaly-free model by requiring that the total WIs are vanishing, independent of $w$ and $z$, as long as they are chosen the same for all fermions in the UV-complete model.  In other words, an anomaly-free model is insensitive to the ambiguity introduced by the momentum shift intrinsic to dimensional regularization, which is a gauge-invariant regularization prescription, as long as the momentum shift is applied consistently for all fermions.

In an effective theory where chiral fermions are taken heavy, the choice of $w$ and $z$ to parametrize the momentum shift in~\eqnsref{ME1}{ME2} also determine the appropriate choice of the Wess-Zumino term~\cite{Wess:1971yu}, which results from the combination of choosing $w$ and $z$ and taking $m \to \infty$ in the WIs.  Since $\lim\limits_{m \to \infty} m^2 C_0 (m) = -1/2$, the heavy fermion mass limit in Eqs.~\ref{eqn:wandzWIs1} and~\ref{eqn:wandzWIs2} exhibits non-decoupling.

Explicitly, we consider the Wess-Zumino term for the hypercharge gauge field $B$, weak gauge fields $W^a$, and a general $Z'$ field,
\begin{align}
\mathcal{L}_\text{WZ} &= C_B g_X g'^2 \epsilon^{\mu \nu \rho \sigma} Z'_{\mu} B_\nu \partial_\rho B_\sigma \nonumber \\ 
&- C_B g_X g^2 \epsilon^{\mu \nu \rho \sigma} Z'_{\mu} \left( W_\nu^a \partial_\rho W_\sigma^a + \frac{1}{3} g \epsilon^{abc} W_\nu^a W_\rho^b W_\sigma^c \right) \ ,
\end{align}
where the coefficient of the weak gauge bosons is negative that of the hypercharge gauge bosons to avoid breaking the electromagnetic gauge symmetry~\cite{DHoker:1984mif, DHoker:1984izu, Dror:2017nsg}.  Isolating the vertex involving $Z-Z'-\gamma$, we get
\begin{align}
    \mathcal{L} \supset -C_B \frac{e_{\text{EM}} g g_X}{c_W} \epsilon^{\mu \nu \rho \sigma} Z'_{\mu} \left( Z_\nu \partial_\rho A_\sigma + A_\nu \partial_\rho Z_\sigma \right) ,
\label{eqn:WZvtx}
\end{align}
which then results in the following WI structure:
\begin{align}
    & ( p_{1 \mu} + p_{2 \mu} ) \Gamma^{\mu \nu \rho} =  C_B \frac{e_{\text{EM}} g g_X}{c_W} \epsilon^{\nu \rho |p_1| |p_2|} \,, 
\label{eqn:cBWIs1} \\
    & -p_{1 \nu} \Gamma^{\mu \nu \rho} = 2 C_B \frac{e_{\text{EM}} g g_X}{c_W} \epsilon^{\mu  \rho |p_1| |p_2|} \,, 
\label{eqn:cBWIs2} \\
    & -p_{2 \rho} \Gamma^{\mu \nu \rho} = C_B\frac{e_{\text{EM}} g g_X}{c_W}\, \epsilon^{\mu \nu |p_1| |p_2|} \,.
\label{eqn:cBWIs3}
\end{align}
We see that these WIs are identical to the contributions of a given heavy fermion with couplings of $g_v^{Z'} = g_a^Z = 0$ and $g_v^Z, g_a^{Z'} \neq 0$ corresponding to the restriction $2z = w-1$ or $g_v^{Z} = g_a^{Z'} = 0$ and $g_v^{Z'}, g_a^Z \neq 0$ corresponding to the restriction $2z = w-3$.  This provides a concrete matching condition for the effective field theory of a heavy chiral fermion in anomalous $U(1)$ gauge theories, and we provide a full discussion on the effective approach in~\appref{comparison}.

In a full theory with a complete, anomaly-free set of fermions, {\it e.g.} for the case of the SM fermions in $U(1)_{B-L}$ or for $U(1)_B$ including anomalons, the dependence on $w$ and $z$ -- and thus the shift dependence -- drops out and the vertex can be calculated unambiguously.  We calculate the induced width for the $Z \to Z' \gamma$ decay for these two complete models in the following subsections~\ref{subsec:U1BL} and~\ref{subsec:U1B}.


\subsection{$U(1)_{B-L}$}
\label{subsec:U1BL}

For $U(1)_{B-L}$ gauge symmetry, the SM fermion content (including three right handed neutrinos) is anomaly free and there is no need to introduce new chiral matter.  The interactions of the $B-L$ gauge boson are given by
\begin{align}
    \mathcal{L}_\text{BL} = g_{BL} \left( \frac{1}{3} \overline{q} \gamma^\mu Z_{BL, \mu}' q - \overline{\ell} \gamma^\mu Z_{BL, \mu}'  \ell - \overline{\nu} \gamma^\mu Z_{BL, \mu}' \nu \right) \ ,
\end{align}
where $q$ includes all up-type and down-type quarks, $\ell$ the charged leptons, and $\nu$ the neutrinos.

\begin{figure}[hb!]
  \includegraphics[width=\linewidth]{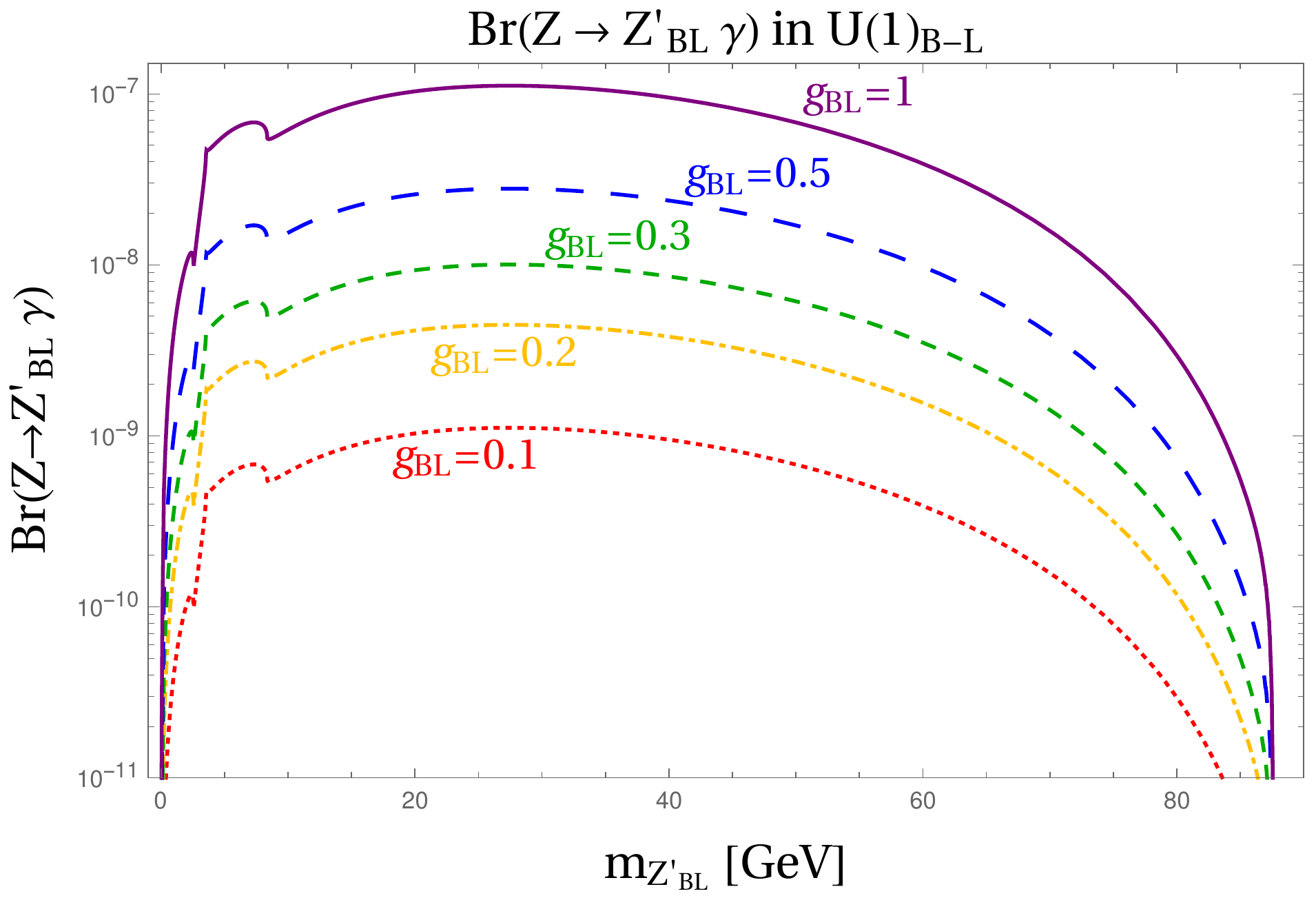}
  \caption{Branching fraction for $Z \to Z_{BL}' \gamma$ for various choices of $g_{BL}$. The top quark dominates the decay width, while the lighter SM fermions cause small peaks when they go on-shell. We include the lighter SM fermion mass effects as in~\eqnref{BLdecay}.}
\label{fig:BRBLF}
\end{figure}

The exotic $Z \to Z'_{BL} \gamma$ decay width, summing over all contributions for the different fermions $f$ with masses $m_f$ and electric charge $Q_f$, is
\begin{widetext}
\begin{align}
    \Gamma (Z \to Z_{BL}^\prime \gamma) = &\dfrac{\alpha_{EM} \alpha \alpha_\text{BL} }{96 \pi^2 c_W^2} \dfrac{m_{Z'}^2}{m_Z} \left( 1 - \frac{m_{Z'}^4}{m_Z^4} \right) \times \nonumber \\
     &\left| \sum\limits_f \left[ T_3^f  N_c^f Q_f^e Q_f^{BL} \left( 2 m_f^2 C_0 (m_f) + \dfrac{m_Z^2}{m_Z^2 - m_{Z'}^2} \left( B_0 (m_Z^2, m_f) - B_0 (m_{Z'}^2, m_f) \right) \right) \right] \right|^2 \ ,
    \label{eqn:BLdecay}
\end{align}
\end{widetext}
where $T_3^f = +1$ for up-type quarks and $-1$ for down-type quarks and charged leptons, $N_c^f = 3$ for quarks and $1$ for leptons, $Q_f^e$ and $Q_f^{BL}$ are the $U(1)$ charges under EM and $B-L$ gauge symmetries, and we also abbreviate
\begin{align}
    B_0(m_V^2, m_f) \equiv B_0( m_V^2, m_f, m_f) \ ,
\end{align}
which is the usual Passarino-Veltman bubble scalar loop function~\cite{Passarino:1978jh}.

From the structure of~\eqnref{BLdecay}, it is clear that a mass-degenerate generation of SM fermions will have a vanishing contribution to the partial width, by virtue of the fact that $\sum_f T_3^f N_c^f Q_f^e Q_f^{BL}$ vanishes for a degenerate set of SM quarks and leptons.  Hence, the largest contribution stems from the intragenerational hierarchy between the top quark and the bottom and tau leptons, as shown in~\figref{BRBLF}. The residual finite mass splittings become more significant for smaller $Z^\prime$ masses: they are visible as bumps in~\figref{BRBLF}. For $m_{Z^{\prime}} \gtrsim 20$~GeV, they have an overall effect of less than $10\%$ compared to keeping them degenerate. Below $m_{Z^{\prime}} \lesssim 10$~GeV, however, the threshold behavior from the lighter fermions can give an enhancement of the decay width by a factor of two compared to the top quark contribution only.

We emphasize that the cancellation of an entire generation of mass-degenerate SM fermions only occurs in this case because the SM fermions share the same underlying chiral symmetry structure which dictates the axial-vector couplings to the $Z$ boson and vector couplings to the $Z'$ boson.  Correspondingly, the expected Landau-Yang behavior~\cite{Landau:1948kw, Yang:1950rg} for $m_{Z'} \to 0$ is also self-evident in~\eqnref{BLdecay} and~\figref{BRBLF}.

\subsection{$U(1)_B$}
\label{subsec:U1B}

For gauged $U(1)_B$ symmetry, with quark interactions as in~\eqnref{U1BLag}, the analytic behavior of $\Gamma(Z \to Z'_B \gamma)$ is markedly different from the $U(1)_{B-L}$ case.  In particular, since the anomaly cancelling fermions can become massive independently of the SM Higgs vev, their contribution to the $Z-Z'-\gamma$ vertex can be non-decoupling regardless of the scale set by the $Z$ mass.  On the other hand, in such a case, the anomalons and the $Z'$ share the same chiral symmetry breaking scale, and the infrared limit of making the $Z'$ light necessarily reintroduces the anomalons into the spectrum too.  The anomalon fields we consider are charged as in~\tableref{quantumnumbers}.

Assuming the charged anomalons are degenerate and their masses $M$ arise solely from $U(1)_B$ breaking, the decay width of $Z \to Z'_B \gamma$ is
\begin{widetext}
\begin{align}
\Gamma(Z \to Z'_B \gamma) &= 
    \dfrac{\alpha_{\text{EM}} \alpha \alpha_X}{96 \pi^2 c_W^2} \dfrac{m_Z'^2}{m_{Z}} \left( 1 - \dfrac{m_{Z'}^4}{m_Z^4} \right) \nonumber \\
    &\Bigg| 
-\sum\limits_{f \in \text{ SM }} T_3(f) Q_f^e
    \left[ \dfrac{m_Z^2}{m_Z^2 - m_{Z'}^2} \left( B_0(m_Z^2, m_f) - B_0(m_{Z'}^2, m_f) \right) + 2 m_f^2 C_0(m_f) \right]
    \nonumber \\
    &
    + 3 \left( \dfrac{m_Z^2}{m_Z^2 - m_{Z'}^2} \left( B_0 (m_Z^2, M) - B_0 (m_{Z'}^2, M)  \right) + 2 M^2 \dfrac{m_Z^2}{m_{Z'}^2} C_0 (M) \right)
\Bigg|^2 \ ,
\label{eqn:ZZbgamma}
\end{align}


\end{widetext}
where again $T_3(f) = +1$ for up-type quarks and $-1$ for down-type quarks.  We remark that it is an excellent approximation (to better than 1\%) to set the masses of the first five flavors of SM fermions as degenerate with  $\mathcal{O}(1)$~MeV masses.  We also remark that, as required by anomaly cancellation, there is no $w$ or $z$ dependence in the physical width.

\begin{figure}
    \begin{subfigure}[t]{0.99\linewidth}
        \includegraphics[width=\linewidth, angle=0]{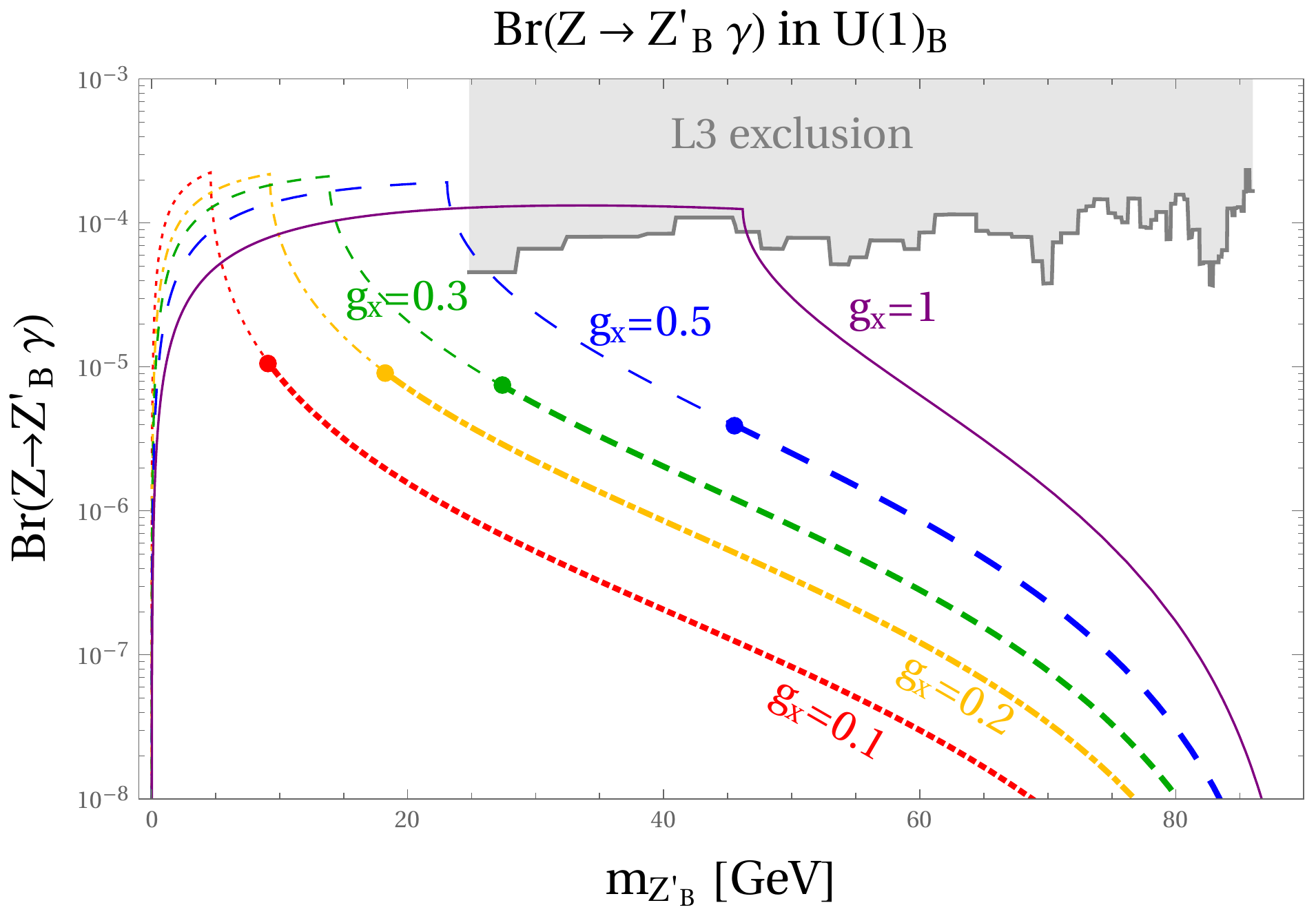}
       \caption{Branching fraction $Z \to Z'_B \gamma$ versus $m_{Z'_B}$.}
       \label{BrZBm}
    \end{subfigure}
    ~\newline\newline
    \begin{subfigure}[t]{0.99\linewidth}
        \includegraphics[width=\linewidth]{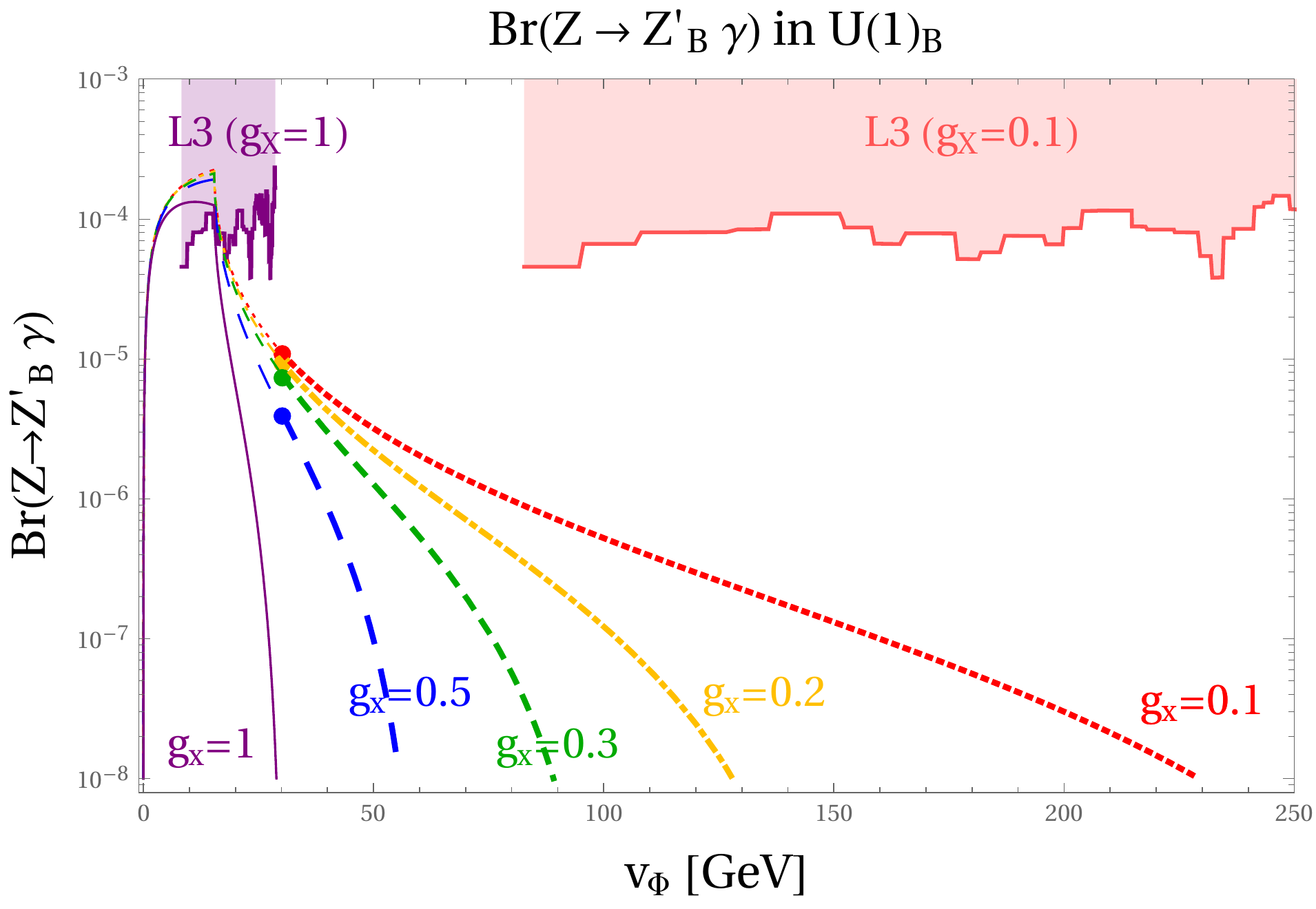}
        \caption{Branching fraction $Z \to Z'_B \gamma$ versus $v_\Phi$.}
        \label{BrZBvev}
    \end{subfigure}
    \caption{Branching fraction for $Z \to Z'_B \gamma$ for various choices of $g_X$. The degenerate anomalon masses $M$ are set to $(4\pi/3) v_\Phi/\sqrt{2}$. The dots mark the point where the anomalon mass equals 90 GeV. The grey (purple, red) shaded region is excluded by the L3 $Z \to (jj)_{\text{res}} + \gamma$ search~\cite{Adriani:1992zm}, and it becomes $g_X$ dependent when plotted versus the vev $v_\Phi$.}
    \label{fig:BRplot}
\end{figure}

We show the branching fraction of $Z \to Z'_B \gamma$ as a function of $m_{Z_B^\prime}$ for various choices of $g_X$ in~\figref{BRplot}. We include the limit on the $Z \to Z_B' \gamma$ branching ratio, where $Z_B'$ decays hadronically, which has been probed at LEP by the L3 collaboration~\cite{Adeva:1991dw, Adriani:1992zm}.  The $Z'_B$ boson will dominantly decay to a dijet resonance for masses $m_{Z'_B}\gtrsim m_{\pi}$~\cite{Dobrescu:2013cmh, Tulin:2014tya}, when the anomalons introduced are heavier than the $Z'_B$ boson.

We remark on many interesting features in~\figref{BRplot}.  First, since the anomaly cancelling fermions obtain mass from the spontaneous breaking of the chiral $U(1)_B$ symmetry, their masses scale with the $U(1)_B$ breaking vev set by $m_{Z_B'} = 3 g_X v_\Phi$.  For concreteness, we set $M = \frac{4\pi}{3} \frac{v_\Phi}{\sqrt{2}}$ using $\frac{4\pi}{3}$ as a fixed value for the Yukawa couplings $y_L$ and $y_E$.  Hence, for fixed $g_X$, the anomalons become lighter as $m_{Z_B'}$ decreases: the cusp behavior at the maximum of each curve then marks when the anomalons develop imaginary contributions to the loop function by going on-shell at $M = m_Z/2$.  Such light, electrically-charged anomalons are already excluded, however, by searches at LEP by the L3 and ALEPH collaborations~\cite{Achard:2001qw, Heister:2002mn}.  Thus, we indicate the LEP direct search bound on the charged anomalons as a solid circle on each curve, marking where their masses cross 90 GeV.  The parameter space for the branching fractions left of these circles, indicated by thinner lines, is thus excluded by direct searches for the charged anomalons.  Of course, the anomalon masses could also receive large contributions from the SM Higgs vev, which would weaken the direct scaling relationship between $m_{Z_B'}$ and $M$ for a given $g_X$, but then the contribution to the anomalon masses from the SM Higgs Yukawa would also affect the $h \to \gamma \gamma$ signal strength.  We evaluate this constraint in~\secref{direct}, finding that anomalons whose dominant mass contribution comes from $v_\Phi$ enjoy an open parameter space to induce a branching fraction of $\mathcal{O}(10^{-5})$.  Finally, the turnover feature of the anomalons is also necessary to exhibit the well-known Landau-Yang behavior~\cite{Landau:1948kw, Yang:1950rg} as $m_{Z_B^{\prime}} \to 0$.

Applying the L3 and ALEPH constraints~\cite{Achard:2001qw, Heister:2002mn} on new electrically charged fermions, the exotic branching of $Z \to Z_B^\prime \gamma$ is necessarily at most $\mathcal{O}(10^{-5})$.  If the L3 and ALEPH constraints were relaxed,\footnote{For example, see the model considered in Ref.~\cite{Egana-Ugrinovic:2018roi}.} then the exotic $Z \to Z_B^\prime \gamma$ branching fraction maximizes around $\mathcal{O}(\text{few}) \times 10^{-4}$, in competition with the exotic $Z \to Z_B^\prime \gamma$ decay probe by L3~\cite{Adeva:1991dw, Adriani:1992zm}. While the hadronic decays of the relatively light $Z_B^\prime$ are more difficult to reconstruct at the LHC compared to LEP, the immense statistics and the additional coincident feature of the $jj + \gamma$ resonance reconstructing the $Z$ boson make this a promising avenue to probe possible anomalous gauge symmetries at the LHC.  We note a similar sensitivity improvement in the exotic decay of the $Z$ to a leptonically decaying $Z'$ and a photon could also be expected from the LHC experiments, where the current branching fraction limits at the $\mathcal{O}(10^{-5})$ level are set by the OPAL collaboration~\cite{Acton:1991dq}.

Finally, we remark that the expression in~\eqnref{ZZbgamma} is the first possible non-trivial decay width of a massive, neutral gauge boson into two further neutral gauge bosons\footnote{For a discussion of the Landau-Yang theorem and its applications to non-Abelian gauge bosons, see Ref.~\cite{Cacciari:2015ela}.}. We also note that heavy sectors of anomaly free sets of fermions, by virtue of the fact that the mixed electroweak anomaly is carried in two distinct vertices, can give a non-decoupling contribution to the partial width, as noted in Ref.~\cite{Dedes:2012me}.  For illustration, a hypothetical complete set of heavy, mass-degenerate SM fermions and anomalons, where the anomalon masses arise solely by the $U(1)_B$ breaking vev, gives a non-decoupling decay width of
\begin{align}
\Gamma(Z \to Z'_B &\gamma)^\text{non-anom.} = \nonumber \\
    &\dfrac{3\, \alpha_{\text{EM}} \alpha \alpha_X}{32\, \pi^2 c_W^2} \dfrac{(m_Z^2 - m_{Z'}^2)^2}{m_Z m_{Z'}^2} \left( 1 - \dfrac{m_{Z'}^4}{m_Z^4} \right) \ .
\end{align}
Of course, the SM-like nature of the 125 GeV Higgs precludes this scenario, but it is nevertheless a curious fact that the decoupling of anomaly free sets is not guaranteed in theories with two sources of chiral symmetry breaking.

\section{Collider searches for anomalons}
\label{sec:direct}

In this section, we discuss the phenomenology of the anomalon sector. Since the anomalons have the same SM gauge quantum numbers as leptons, they share many of the same phenomenological signatures as fourth-generation leptons.  Given their dominant mass contribution arises from $v_\Phi$, their collider phenomenology also mimics the electroweakino sector from supersymmetry, where the charged anomalons are slightly heavier than the electrically neutral anomalons and exhibit a compressed mass spectrum, as a consequence of the 1-loop radiative electroweak corrections. 

When the anomalon masses receive contributions from the electroweak vev, they induce corrections to the observed 125~GeV Higgs boson decay into two photons.  We can calculate this correction as a coherent sum of the top quark, bottom quark, $W$ boson, and the new anomalons.  Since we assume the dominant source of the anomalon masses comes from $v_\Phi$, they will exhibit decoupling in the $H \to \gamma \gamma$ partial width.

From~\eqnref{Lyuk}, we assume the Yukawa couplings are real and for simplicity set $y_E = y_L \equiv y_\Phi$ and $y_1 = y_2 \equiv y_H$.  The charged anomalon mass Lagrangian becomes 
\begin{align}
\mathcal{L}_{\text{mass}} \supset -\frac{y_{\Phi}}{\sqrt{2}} v_{\Phi }\bar{e}_L e_R - \frac{y_{\Phi}}{\sqrt{2}} v_\Phi  \bar{E}_L E_R\nonumber \\- \frac{y_H}{\sqrt{2}} v_H \bar{E}_L e_R - \frac{y_H}{ \sqrt{2}} v_H \bar{e}_L E_R + \text{ h.c.} \,,
\end{align}
and the two Dirac masses are then
\begin{align}
M_1 = \frac{1}{\sqrt{2}} \left( y_\Phi v_\Phi + y_H v_H \right) \ , \\
M_2 = \frac{1}{\sqrt{2}} \left| y_\Phi v_\Phi - y_H v_H \right| \ .
\end{align}
The mass eigenstates couple to the SM Higgs with the Yukawa coupling $\pm y_H$, while the Dirac masses depend on the values of $y_H$, $y_\Phi$ and the vev $v_\Phi$.

Adapting the expression for the Higgs decay to two photons from Ref.~\cite{Djouadi:2005gi}, the $H \to \gamma \gamma$ partial width including the two additional charged fermions becomes
\begin{align}
    \Gamma (H\to \gamma \gamma) = &\frac{G_F \alpha_{\text{EM}}^2 M_H^3}{128 \sqrt{2} \pi^3} 
    \left | \frac{4}{3} A_f(\tau_t) + \frac{1}{3} A_f(\tau_b) + A_W (\tau_W)
    \right.\\ 
    &\left. +\frac{y_H v_H}{M_1 \sqrt{2}} A_f (\tau_{M1}) -\frac{y_H v_H}{M_2 \sqrt{2}} A_f (\tau_{M2}) \right|^2 \ , \nonumber 
\end{align}
where $G_F$ is the Fermi constant, $M_H$ is the Higgs mass of 125~GeV, and we include the dominant contributions from the top quark, bottom quark, the $W$ boson, and the two new fermions.  The loop functions are defined as
\begin{eqnarray}
A_f(\tau_f)&=& \frac{2}{\tau_f^2} (\tau_f+(\tau_f-1)f(\tau_f)) \, ,\\
A_W(\tau_W)&=& -\frac{1}{\tau_W^2} (2\tau_W^2+3\tau_W+3 (2\tau_W-1)f(\tau_W))\, ,
\end{eqnarray}
with the function $f(\tau)$ being
\begin{equation}
  f(\tau)=\begin{cases}
    (\arcsin{\sqrt{\tau}})^2, & \tau \leq 1\\
    -\frac{1}{4}\left( \log{\frac{1+\sqrt{1-\tau^{-1}}}{1-\sqrt{1-\tau^{-1}}}} -i\pi \right)^2, & \tau > 1
  \end{cases}
\end{equation}
and the arguments $\tau$ are the mass ratios $\tau_{f,W}=\frac{M_H^2}{4 m_{f,W}^2}$.

\begin{figure}
    \centering
        \includegraphics[width=0.45\textwidth]{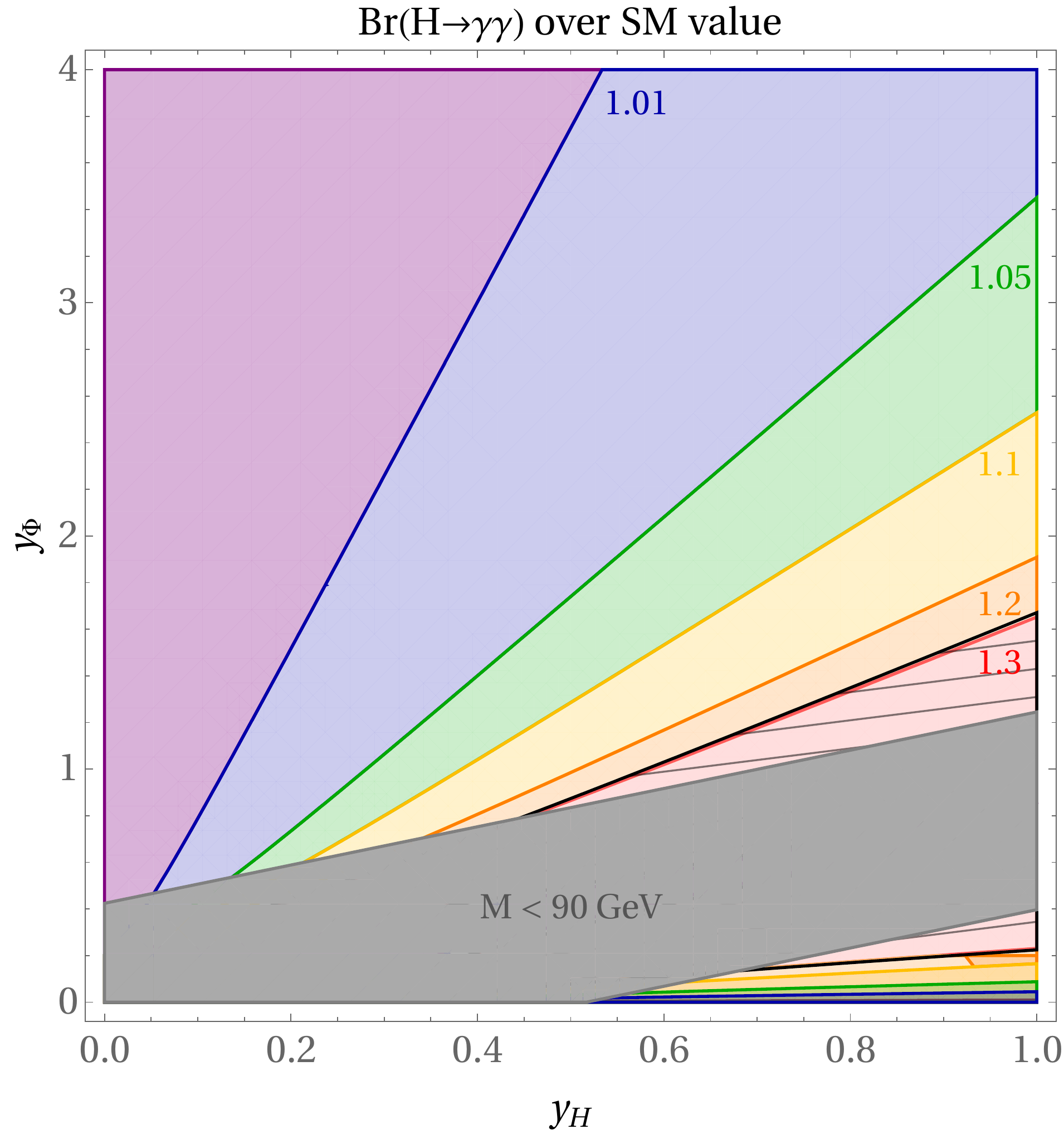}
       \caption{$H\rightarrow \gamma \gamma$ branching fraction exclusion plot including both charged anomalons, where $y_H$ is the coupling to the SM Higgs boson and $y_{\Phi}$ the one to the $U(1)_B$ breaking Higgs. The vev of the new Higgs is fixed to $v_\Phi=300$ GeV. The dashed regions are the $H \to \gamma \gamma$ signal strength $2\sigma$ exclusion limits~\cite{Aaboud:2018xdt}. The grey band denotes the 90 GeV exclusion limit on the anomalon masses by LEP~\cite{Achard:2001qw, Heister:2002mn}.}
       \label{fig:yHyPhihggexclusion}
\end{figure}

In~\figref{yHyPhihggexclusion}, we plot contours of the signal strength for $gg \to H \to \gamma \gamma$ as a function of the two Yukawa couplings of the anomalons, $(y_H, y_\Phi)$ where $v_\Phi = 300$~GeV.  The open region shows the parameter space allowed by the ATLAS limits~\cite{Aaboud:2018xdt} on the signal strength of
\begin{align}
    \mu (gg \to H \to \gamma \gamma) = 0.99^{+0.15}_{-0.14} \ ,
\end{align}
where the $1\sigma$ uncertainties are given.  The hatched grey regions in~\figref{yHyPhihggexclusion} are the two sigma exclusion limits.  We also show the exclusion from LEP searches on charged particles below $90$~GeV~\cite{Achard:2001qw, Heister:2002mn}.  We see that the charged anomalons can have a mild effect on the $H \to \gamma \gamma$ rate, which is well within the experimental uncertainty when $y_\Phi$ is dominant over $y_H$.  This is a direct result of their vector-like SM gauge representations.  As a result, we expect the best improvement in testing this parameter space would come from direct searches for anomalons, subject to the model dependence in their decay signatures. Note that there is also a region at very small $y_\Phi$ but high $y_H$ that is allowed by the $H \to \gamma \gamma$ rate, but we expect this to be excluded by electroweak precision measurements.

In our model, if the electrically neutral anomalons $\nu$ and $N$ are only slightly lighter than the electrically charged anomalons, we would have a completely analogous situation to the electroweakino and slepton searches from supersymmetry, which are one of the more difficult signatures for the LHC experiments because of the presence of soft leptons from compressed mass splittings~\cite{Schwaller:2013baa, Aad:2019vnb, Aad:2019qnd}. The $SU(2)$ couplings of the anomalons would guarantee electroweak Drell-Yan production rates, but for small mass splittings, the charged current decay from the heavy charged anomalon to the electrically neutral anomalon would give very soft leptons or pions and missing transverse energy.  As shown in Ref.~\cite{Aad:2019qnd}, the most pessimistic choice of mass splitting means there is no limit from the LHC experiments, and we can only adopt the limit from the LEP searches.  We reserve a dedicated study of the collider phenomenology of direct anomalon searches and the suitability of the lightest neutral anomalon as a dark matter candidate for future work.

\section{Coupling-mass Mapping for $U(1)_B$}
\label{sec:couplingmass}

\begin{figure}[tb]
\begin{center}
\includegraphics[width=\linewidth, angle=0]{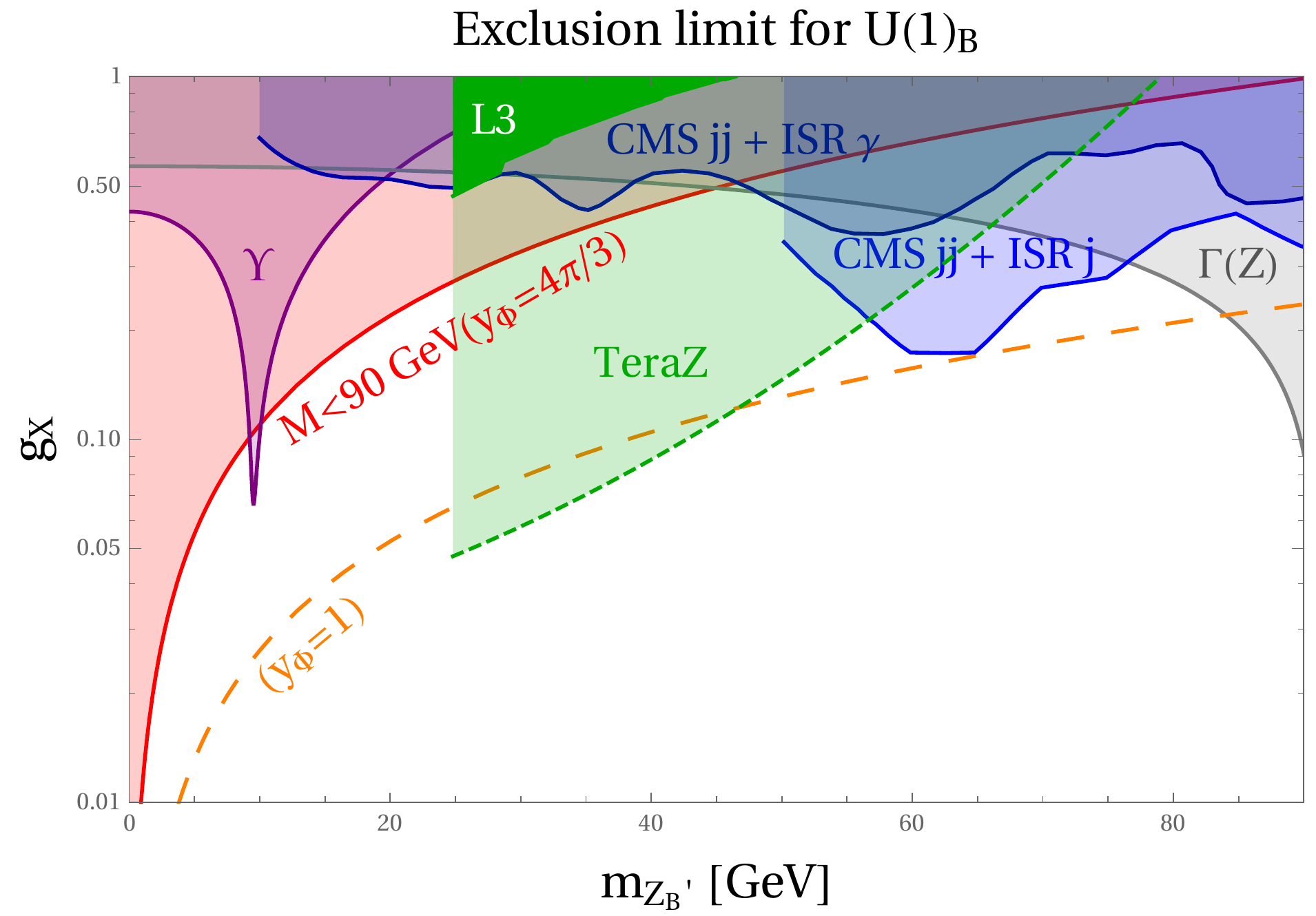}
\caption{Exclusion limits for $U(1)_B$ in the $g_X$, $m_{Z_B'}$ plane including the limit from the decay of the $\Upsilon$ (in purple) and the limit from the width of the $Z$ boson (in gray), both taken from Ref.~\cite{Dobrescu:2014fca}, our limit from $Z \to jj \gamma$ data from the L3 experiment at LEP~\cite{Adriani:1992zm} (in solid green) and a TeraZ factory projection in lighter green. The direct bound on the anomalon mass~\cite{Achard:2001qw, Heister:2002mn} is plotted for two different maximum values for the Yukawa coupling, $y_\Phi = 1$ (in orange, dashed) and $y_\Phi = 4\pi/3$ (in red). The low-mass dijet resonance limit from CMS based on a dijet resonance + ISR photon search~\cite{Sirunyan:2019sgo} is shown in darker blue, while the jet substructure + ISR jet search~\cite{Sirunyan:2019vxa} is shown in lighter blue.}
\label{fig:gX_exc}
\end{center}
\end{figure}

We now discuss the overall status of $Z'_B$ boson searches in the mass region accessible by the $Z \to Z'_B \gamma$ decay.  Our summary plot is shown in~\figref{gX_exc}, with numerous searches carving out excluded regions in the $g_X$ vs.~$m_{Z'_B}$ plane. 

The first constraint in~\figref{gX_exc} is marked ``L3" and is derived from the search for the $Z \to Z'_B \gamma$, $Z'_B \to jj$ exotic decay~\cite{Adriani:1992zm}. It is calculated following~\eqnref{ZZbgamma}, setting the mass of the anomalons by their nominal Yukawa coupling of $4\pi/3$ and using the SM value for the top quark mass. The other SM quark masses are set to be of order MeV. Note that solving the branching fraction at a given mass $m_{Z'}$ for the value of $g_X$ gives two values for $g_X$, where the limit is given by the smaller one. In the case of a discovery, however, this uncertainty has to be resolved by an independent measurement.
We see that the L3 constraint is largely supplanted by the more recent CMS searches targeting low-mass dijet resonances triggered using initial state radiation photons~\cite{Sirunyan:2019sgo} or jets, with a subsequent jet substructure analysis~\cite{Sirunyan:2019vxa}.

In~\figref{gX_exc}, we also show the constraint based on the $\Upsilon$ search by the ARGUS collaboration~\cite{Albrecht:1986ec}, where they measured the hadronic ratio of $\Upsilon$ decays, $R_\Upsilon = \Gamma(\Upsilon \to \text{   hadrons})/\Gamma(\Upsilon \to \mu^+ \mu^-) < 2.1$.  The corresponding limit in $g_X$ vs.~$m_{Z'}$ is taken from Refs.~\cite{Aranda:1998fr, Dobrescu:2014fca}.  We also take the limit from Ref.~\cite{Dobrescu:2014fca} for $Z-Z'_B$ mixing induced at one-loop by SM quarks~\cite{Carone:1994aa, Bailey:1994qv, Carone:1995pu, Aranda:1998fr, Dobrescu:2014fca}.  This constraint is labeled $\Gamma(Z)$ in~\figref{gX_exc}.

The LEP constraint from the L3 and ALEPH collaborations that new electrically charged particles must be heavier than at least 90~GeV~\cite{Achard:2001qw, Heister:2002mn} is also shown in~\figref{gX_exc} for $y_\Phi = 4\pi/3$, which is the same value of $y_\Phi$ that we use for the anomalon masses in deriving the L3 limit.  We also show the LEP constraint for a different choice of $y_\Phi = 1$ as a dashed line.  We remark that these constraints are subject to additional model dependence, since these curves have also set $y_H = 0$.  Importantly, as shown in~\figref{yHyPhihggexclusion}, we can turn on nonzero $y_H$ to increase the anomalon masses, which would make the collider dijet searches, the $\Upsilon$ bound, and the $Z-Z'_B$ mixing bound become the leading existing bounds on these light $Z'_B$ bosons.

Finally, we also show a projection for a TeraZ factory searching for the exotic $Z \to Z'_B \gamma$, $Z'_B \to jj$ decay as a dotted green outlined region in~\figref{gX_exc}.  For this projection, we perform a simple extrapolation accounting for the increased statistics of $3 \times 10^{12}$ $Z$ bosons produced during four years of runtime of the FCC-ee~\cite{Blondel:2018mad} over the number of $Z$ bosons produced at LEP, which was $1.7\times 10^{7}$~\cite{ALEPH:2005ab}.  We also simply assume the improvement in sensitivity only arises from the increased statistics, but we recognize that a realistic collider projection should also account for possible improvements in mass reach as well as problematic background or systematic uncertainties.  We reserve a such a sensitivity study for future work.

Importantly, the exotic $Z \to Z'_B \gamma$ decay is an irreducible probe of the $U(1)_B$ model compared to the anomalon searches, which depend on the specific decay channels employed by the anomalons.  The $Z \to Z'_B \gamma$ rate also crucially retains information about the anomaly coefficient contribution from the heavy fermions, which matches the anomaly coefficient of the SM fermion content.  In fact, the gauge coupling and $Z'$ mass are the only continuous parameters that dictate the exotic $Z \to Z' \gamma$ rate, since the nonzero anomaly coefficients for different global $U(1)$ symmetries of the SM fermions are necessarily discrete.  Hence, future improvements on the $Z \to Z_B' \gamma$ exotic decay search are strongly motivated.

\section{Conclusions}
\label{sec:conclusions}

In this work, we have presented the calculation and phenomenology of the exotic $Z$ decay to a $Z'$ boson and a photon.  Our calculation, while reminiscent of textbook calculations of anomaly coefficients, focuses on the full $Z-Z'-\gamma$ vertex function, where the various Ward-Takahashi identities are obtained by taking the appropriate divergences of the external currents.  We illustrated the different types of anomaly contributions and non-decoupling behavior in the Ward-Takahashi identities by considering two distinct $U(1)$ gauge symmetries: $U(1)_{B-L}$ gauge symmetry and $U(1)_B$ gauge symmetry.  In the $B-L$ case, since the SM fermion content is anomaly-free, they necessarily have vector-like representations under $U(1)_{B-L}$ symmetry.  Moreover, a degenerate set of anomaly-free fermions would have a vanishing contribution to the $Z \to Z'_{BL} \gamma$ decay width.

In contrast, the SM fermions give non-vanishing electroweak anomalies when gauging baryon number.  Correspondingly, when the anomalon masses arise only from $U(1)_B$ breaking, the anomaly cancellation between SM fermions and anomalons ensures the exotic $Z \to Z'_B \gamma$ decay width does not vanish if all fermions are degenerate.  This feature, which was also noted in Ref.~\cite{Dedes:2012me}, is directly related to the fact that two sets of fermions can have distinct chiral symmetry breaking scales.  As a result, the phenomenological predictions from $U(1)_B$, including the exotic $Z \to Z'_B \gamma$ decay, searches for the electroweak charged anomalons at colliders, low-mass $Z'_B$ dijet resonance searches, and shifts in $Z$-pole observables from $Z-Z'_B$ mixing, are all different manifestations of the underlying scale of $U(1)_B$ breaking for their respective particle content.  The corresponding state of the art, shown in~\figref{gX_exc}, shows that all of these diverse probes with their separate considerations of backgrounds and systematic uncertainties are nonetheless generally competitive.

We also dedicated~\appref{comparison} to the discussion of the non-decoupling nature of the anomalon fields and the corresponding induced Wess-Zumino term as well as a discussion of the Goldstone boson equivalence treatment of the $Z \to Z' \gamma$ decay.  This Appendix demonstrates that an ultraviolet complete theory predicts markedly different behavior for the $Z \to Z' \gamma$ decay compared to any approach that neglects the SM fermions, especially for realistic new $U(1)$ gauge couplings that can be probed at current and near future colliders.

We see that the phenomenology of new $U(1)$ symmetries is exceedingly rich in both the theoretical complexity and the phenomenological predictions.  We have highlighted the particularly special behavior of anomalous $U(1)$ symmetries with the $U(1)_B$ case study and the non-decoupling of anomalons in the corresponding $Z-Z'_B-\gamma$ vertex.  We find that the $Z \to Z'_B \gamma$ decay can prove an exciting test of the $U(1)_B$ model when this exotic decay is kinematically accessible. 

\section*{Acknowledgments}
\label{sec:acknowledgments}

FY would like to acknowledge helpful discussions with Bogdan Dobrescu, Andrey Katz, and Toby Opferkuch.  The authors would like to thank Joachim Kopp for helpful discussions and suggestions on the manuscript.  This research is supported by the Cluster of Excellence ``Precision Physics, Fundamental Interactions and Structure of Matter" (PRISMA$^{+}$-EXC 2118/1). The work of LM is also supported by the German Research Foundation (DFG) under Grants No. \mbox{KO 4820/1--1}, and No. FOR 2239, and from the European Research Council (ERC) under the European Union’s Horizon 2020 research and innovation program (Grant No. 637506, ``$\nu$Directions"). Feynman diagrams were generated using jaxodraw~\cite{Binosi:2008ig}.

\begin{appendix}
\section{Effective operator treatment for $Z \to Z'_B \gamma$}
\label{sec:comparison}

In this appendix, we perform an effective operator treatment of our $Z \to Z'_B \gamma$ exotic decay calculation.  In particular, we compare to the calculation performed in Refs.~\cite{Dror:2017ehi, Dror:2017nsg}, where the authors show a result based on an effective operator inducing an enhanced longitudinal coupling to the $Z'_B$ boson.  We will derive the effective operator from our matrix elements in~\eqnsref{ME1}{ME2}, reiterate the discussion about the resulting Ward-Takahashi identities in~\eqnref{wandzWIs1}$-$~\eqnref{wandzWIs3}, and analyze the validity of assuming longitudinal dominance.

We first outline the effective ansatz being made in Refs.~\cite{Dror:2017ehi, Dror:2017nsg}.  The calculation starts by integrating out the anomaly cancelling physics and replacing their effects by Wess-Zumino terms in the Lagrangian,
\begin{align}
\mathcal{L}_{\text{WZ}}&=C_B g_X g'^2 \epsilon^{\mu\nu\rho\sigma} Z'_{\mu} B_{\nu} \partial_{\rho} B_{\sigma}
\label{eqn:CB_WZterms} \\ 
\nonumber &- C_B g_X g^2 \epsilon^{\mu\nu\rho\sigma} Z'_{\mu} (W_{\nu}^a \partial_{\rho} W_{\sigma}^a + \frac{1}{3} g \epsilon^{abc} W_{\nu}^a W_{\rho}^b W_{\sigma}^c )\,,
\end{align}
where $C_B$ is the Wilson coefficient from decoupling the anomalons.  We remark that the Yukawa couplings of the chiral anomalons cannot be arbitrarily large, and thus fixing the mass of the $Z'_B$ boson and the gauge coupling $g_X$ will also establish an upper bound on the perturbative masses of the anomalons.  Hence, for a given combination of $m_{Z'_B}$ and $g_X$, keeping the anomalons decoupled can be inconsistent: in fact, it is responsible for the peak behavior already seen in~\figref{BRplot} and recovering the Landau-Yang limit for $m_{Z'_B} \to 0$, which is the regime of validity motivated by the Goldstone boson equivalence approach.

Refs.~\cite{Dror:2017ehi, Dror:2017nsg} proceed with the Wess-Zumino terms in~\eqnref{CB_WZterms} by using the Goldstone boson equivalence (GBE) theorem to calculate the longitudinally enhanced parts of the amplitude. In~\subsecref{vertex} we deduced the contribution of the Wess-Zumino Lagrangian to the $Z$--$Z'$--$\gamma$ vertex in~\eqnref{WZvtx}. By replacing the $Z'$ boson with the derivatively coupled, linearly realized Goldstone pseudoscalar, $Z'_{\mu} \rightarrow \partial_\mu \varphi / (g_X f_{Z'})$, we obtain the longitudinally equivalent Lagrangian,
\begin{align}
\mathcal{L} &\supset C_B \frac{\varphi}{f_{Z'}} \cdot gg'Z_{\mu\nu}\tilde{F}^{\mu\nu} \, ,
\label{eqn:GBELang}
\end{align}
after an integration by parts, where $f_{Z'}=m_{Z_B'}/g_X$ is the pseudoscalar decay constant, $\tilde{F}^{\mu\nu}=(1/2) \, \epsilon^{\mu\nu\rho\sigma}F_{\rho\sigma}$ is the dual electromagnetic field tensor, $C_B = \mathcal{A}_{Z'BB} / (16 \pi^2)$ and $\mathcal{A}_{Z'BB} = -3/2$ is the baryon number anomaly coefficient from SM fermions~\cite{FileviezPerez:2010gw}.
The GBE is valid in the limit $m_{Z'}$ and $g_X$ each go to zero, with their ratio remaining constant.  By construction, since the GBE also neglects the transverse modes of the $Z'_B$, this treatment breaks down as the $Z'_B$ mass grows.

Given the interaction in~\eqnref{GBELang}, we now can calculate the exotic width of $Z \to \varphi \gamma$:
\begin{align}
    \Gamma_{GBE} &= \frac{1}{24\pi} C_B^2 \frac{e_{\text{EM}}^2 g^2}{c_W^2} \frac{m_Z^3}{f_{Z'}^2} \left( 1 - \frac{m_{\varphi}^2}{m_Z^2} \right)^3 \nonumber \\
    &\approx 
    \frac{1}{24\pi} C_B^2 \frac{e_{\text{EM}}^2 g^2}{c_W^2} \frac{m_Z^3}{f_{Z'}^2} \nonumber \\
    &= \frac{3 e_{\text{EM}}^2 g^2 g_X^2}{8192 \pi^5 c_W^2} \frac{m_Z^3}{m_{Z'}^2} \ ,
\label{eqn:GBEwidth}
\end{align}
where the second line approximates the Goldstone mass small compared to the $Z$ mass and then we used $\mathcal{A}_{Z'BB} = -3/2$.

\subsection{Procedure for reproducing the Goldstone boson equivalence ansatz from the full vertex calculation}

Having derived the GBE result in~\eqnref{GBEwidth}, we critically evaluate the simplifications needed to reduce our full result in~\eqnref{ZZbgamma} to the approximation above.  In sequence, these simplifications are
\begin{enumerate}
    \item ``Integrate out" the anomalon field content and substitute their loop contribution by a Wess-Zumino term.
    \item Approximate all SM fermion masses, including the top quark mass, as negligible compared to the $Z$ and $Z'_B$ masses.
    \item Neglect the transverse mode of the $Z'_B$ gauge boson in the sum over polarization vectors.
\end{enumerate}

{\bf Step 1:} We start with the WIs in Eqs.~(\ref{eqn:wandzWIs1})$-$(\ref{eqn:wandzWIs3}) and consider anomalons with pure vector couplings to the $Z$ boson and axial-vector couplings to the $Z'_B$ boson.  Taking $\lim\limits_{m \to \infty} m^2 C_0 (m) = -1/2$ and summing over the anomalon fields in~\tableref{quantumnumbers}, we find the net anomalon WI contributions are\footnote{In the case that the anomalons are SM-like in their Yukawa couplings, we have $(w - z - 2)$, $(w - 1)$ and $(z+1)$ as the coefficients instead.}
\begin{align}
    \left( p_{1 \mu} + p_{2 \mu} \right) \Gamma^{\mu \nu \rho} &= \frac{3 e_{\text{EM}} g g_X}{16 \pi^2 c_W} \epsilon^{\nu \rho |p_1| |p_2|} (w-z) \,,
    \label{eqn:wandzWIs1_heavy}
    \\
    -p_{1 \nu} \Gamma^{\mu \nu \rho} &= 
    \frac{3 e_{\text{EM}} g g_X}{16 \pi^2 c_W} \epsilon^{\mu \rho |p_1| |p_2|} (w+1) \,,
    \label{eqn:wandzWIs2_heavy}
    \\
    -p_{2 \rho} \Gamma^{\mu \nu \rho} &= 
    \frac{3 e_{\text{EM}} g g_X}{16 \pi^2 c_W} \epsilon^{\mu \nu |p_1| |p_2|} (z+1) \, .
    \label{eqn:wandzWIs3_heavy}
\end{align}
As discussed below Eqs.~(\ref{eqn:cBWIs1})$-$(\ref{eqn:cBWIs3}), introducing a Wess-Zumino effective interaction in order to integrate out the anomalons dictates a specific choice of $w$ and $z$.  

In this case, with the anomalon mass eigenstates having $g_a^Z = 0$, we need $2z = w-1$ to match the Wess-Zumino interaction contributions to the WIs.  The matching condition requires $C_B = 3(z+1) / (16 \pi^2)$, with $C_B$ given in the Wess-Zumino term~\eqnref{WZvtx}.  This discussion makes it manifest that a particular Wess-Zumino term is not independent of the momentum shift relevant for the SM fermions, which remain in the effective theory description.

{\bf Step 2:} Correspondingly, the SM fermions give 
\begin{align}
    \left( p_{1 \mu} + p_{2 \mu} \right) \Gamma^{\mu \nu \rho} &= -\frac{3 e_{\text{EM}} g g_X}{16 \pi^2 c_W} \epsilon^{\nu \rho |p_1| |p_2|} (w-z) \,,
    \label{eqn:wandzWIs1_SM}
    \\
    -p_{1 \nu} \Gamma^{\mu \nu \rho} &= 
    -\frac{3 e_{\text{EM}} g g_X}{16 \pi^2 c_W} \epsilon^{\mu \rho |p_1| |p_2|} (w-1) \,,
    \label{eqn:wandzWIs2_SM}
    \\
    -p_{2 \rho} \Gamma^{\mu \nu \rho} &= 
    -\frac{3 e_{\text{EM}} g g_X}{16 \pi^2 c_W} \epsilon^{\mu \nu |p_1| |p_2|} (z+1) \, ,
    \label{eqn:wandzWIs3_SM}
\end{align}
where we have set all SM fermions to be massless.  We remark that the sum of SM fermion and anomalon contributions to the Ward identities to the photon vertex always cancels, regardless of the choice of the shift parameter $z$.   Summing over the WZ term and SM fermion contributions to the WIs, the only non-vanishing WI is from the $Z'_B$ boson, as expected.

{\bf Step 3:}
After evaluating the matrix elements in~\eqnsref{ME1}{ME2}, setting all SM fermion masses to zero and all anomalon masses to infinity, we next consider the sum over final state polarizations of the $Z'_B$ boson.  In the typical sum, $\sum \epsilon_\nu (p_1) \epsilon_\beta^* (p_1) = -g_{\nu \beta} + \frac{p_{1 \nu} p_{1 \beta}}{m_{Z'}^2}$, we can discard the transverse component of the $Z'_B$ boson by neglecting the $-g_{\nu \beta}$ term.  Moreover, the $Z'_B$ momentum coupling that remains is exactly the derivatively-coupled Goldstone boson that leads to the Lagrangian term in~\eqnref{GBELang}.

\begin{figure*}[htb!]
    \begin{subfigure}[t]{0.49\textwidth}
        \includegraphics[width=\textwidth]{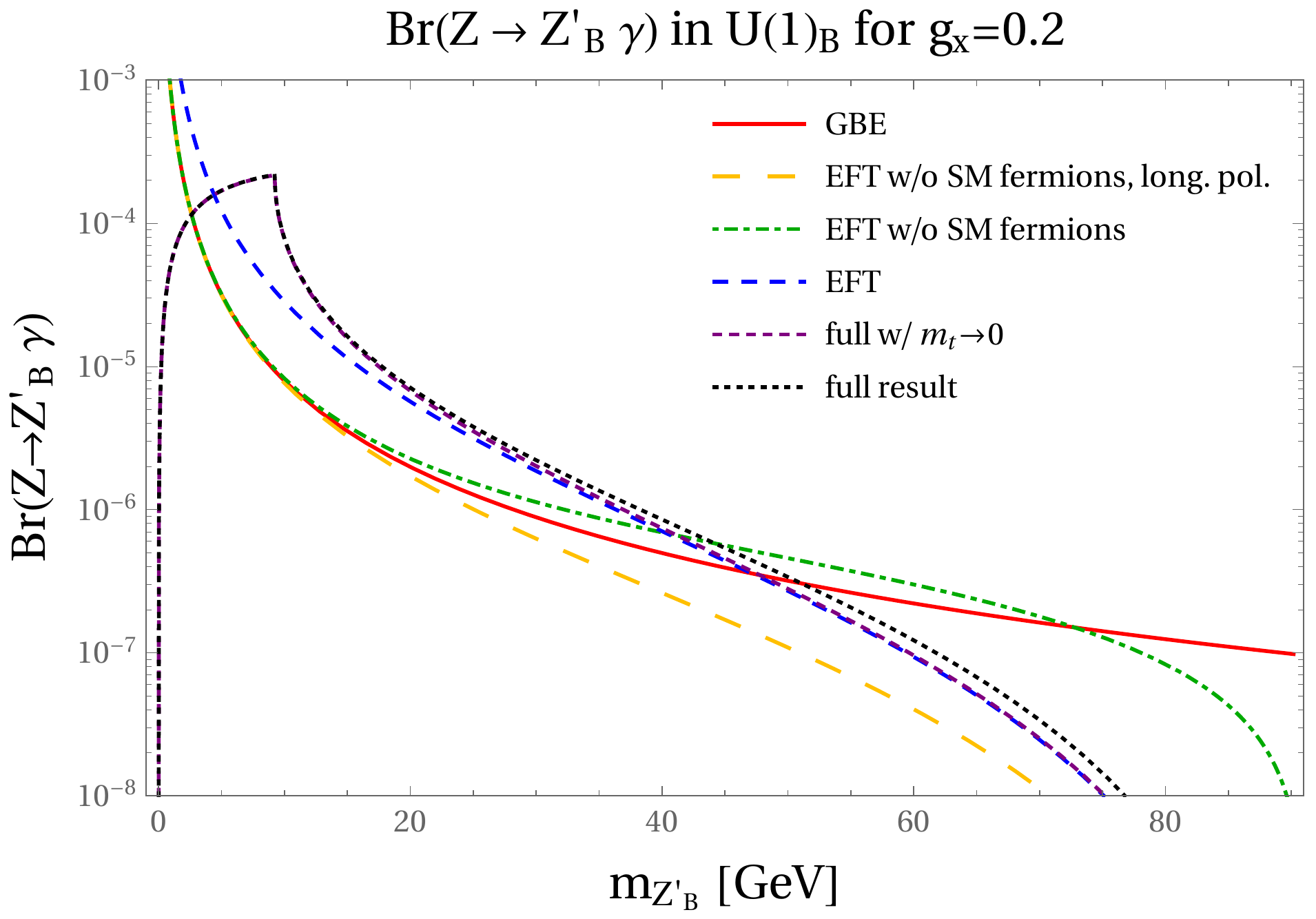}
       \caption{Branching fraction $Z \to Z_B' \gamma$ against $m_{Z'}$ for $g_X=0.2$.}
       \label{BrPosp08}
    \end{subfigure}
    \begin{subfigure}[t]{0.49\linewidth}
        \includegraphics[width=\textwidth]{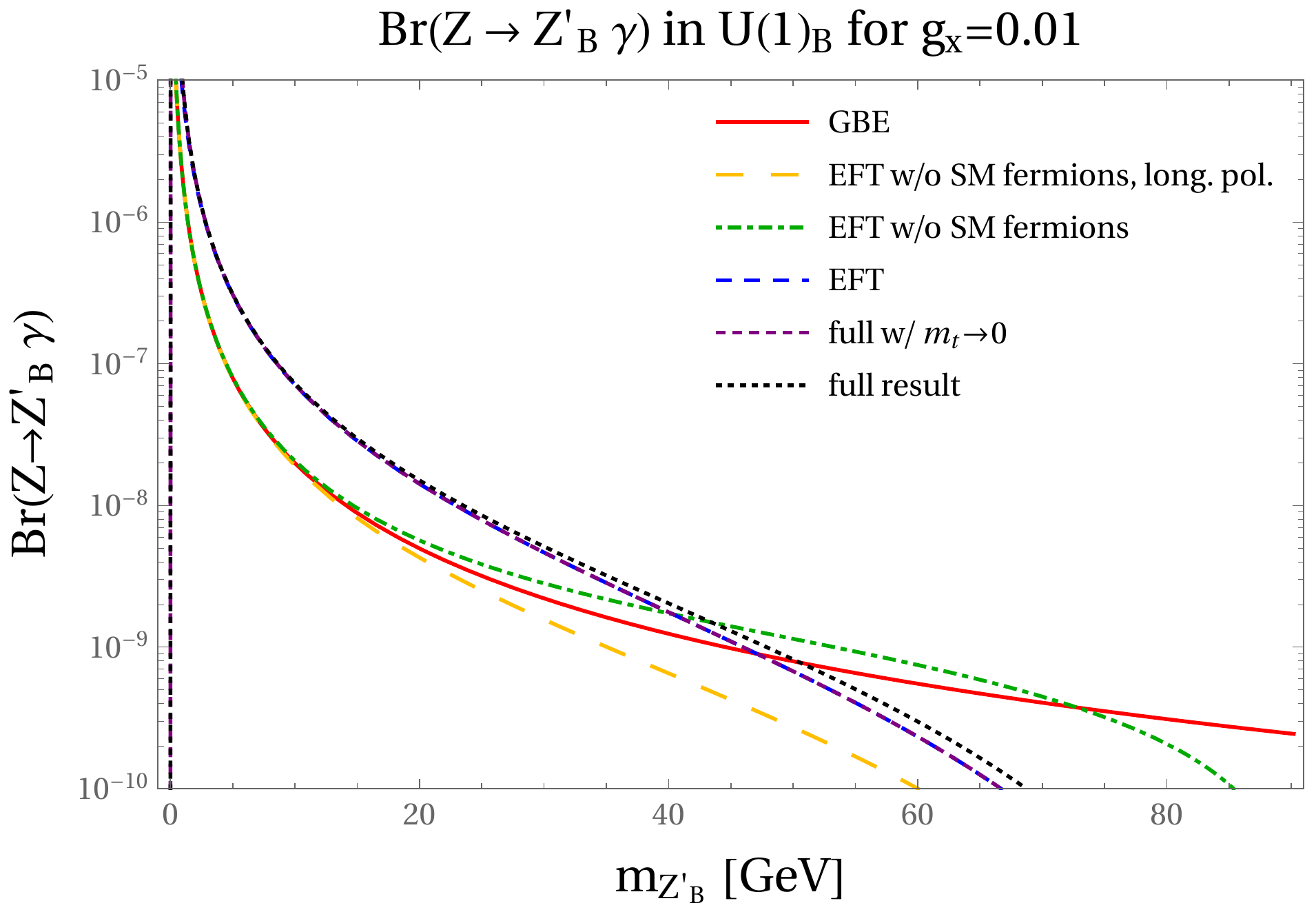}
        \caption{Branching fraction $Z \to Z_B' \gamma$ against $m_{Z'}$ for $g_X=0.01$.}
        \label{BrPosp001}
    \end{subfigure}
    \caption{Branching fraction for $Z \to Z_B' \gamma$ for two values of $g_X$. The red curve takes the width from GBE, see~\eqnref{GBEwidth}. It is calculated from a WZ term and neglects SM fermions, the transverse polarisation of the $Z_B'$ and approximates $m_Z \gg m_{Z'}$. The curve ``EFT w/o SM fermions, long. pol." (yellow, large dashing) is an effective approximation in our calculation, where we include anomalons only with $M \to \infty$ and we thus have to set $w$ and $z$ as explained in ``Step 1" in the text, and we take the longitudinal polarisation only. ``EFT w/o SM fermions" (green, dot-dashed) adds the full polarisation hereto. In ``EFT" (blue, medium dashed) we include the SM quarks (with masses $m_q \to 0$) and anomalons (with masses $M \to \infty$). The curve ``full w/ $m_t \to 0$" (purple, small dashing) sets the anomalon Yukawa couplings to a maximum value of $4\pi/3$ but all SM fermion masses including the top to order MeV. The ``full result" (black, dotted) takes anomalon Yukawas of $4\pi/3$, the SM value for $m_t$ and other SM fermion masses of order MeV.}
    \label{fig:Br_Pospelov}
\end{figure*}

The resulting longitudinal decay width, where we also expand in powers of $m_{Z'}^2 / m_Z^2$, is
\begin{align}
    \Gamma (Z \to Z'_{B, \text{ long}} \gamma) = \frac{3 e_{\text{EM}}^2 g^2 g_X^2 m_Z^3}{2048 \pi^5 c_W^2 m_{Z'}^2} \left( 1 + \mathcal{O} \left( \frac{m_{Z'}^2}{m_Z^2} \right) \right)^3 \ .
    \label{eqn:ZBlong}
\end{align}
We see that our longitudinal width in~\eqnref{ZBlong}, which takes into account the massless SM fermions as well as the heavy anomalons, is a factor of 4 larger than the result obtained by the GBE assumption in~\eqnref{GBEwidth}.  Crucially, our procedure does reproduce the GBE-derived width only when the SM fermions are completely neglected.  In that case, the momentum shift parameters are fixed to the relations specified above to appropriately match to a Wess-Zumino operator description and the normalization set by the anomaly coefficient $\mathcal{A}_{Z'BB} = -3/2$.  In addition, the longitudinal width in~\eqnref{ZBlong} is also the limiting case of our full result in~\eqnref{ZZbgamma} where the SM fermions are massless, the anomalon masses are decoupled, and we perform a series expansion for small $m_{Z'}^2 / m_Z^2$.  Thus, we have established that the GBE width is an incorrect result for the exotic decay of $Z \to Z'_B \gamma$, if the SM fermions are neglected in the calculation.

We show a comparison between the GBE result and our full result after applying successive approximations in~\figref{Br_Pospelov}, including illustrative intermediate effective theory results.

Having explained the different steps from the full calculation,~\eqnref{ZZbgamma}, to a GBE ansatz,~\eqnref{GBEwidth}, we want to illustrate the impact of the different approximations in~\figref{Br_Pospelov}. It shows the branching fraction of $Z \to Z_B' \gamma$ for a rather large value of $g_X=0.2$ and for smaller $g_X=0.01$. We want to begin the discussion from the ``full result" (black, dotted curve). It takes the width from ~\eqnref{ZZbgamma}, setting the anomalon masses to their maximum value of $M=4\pi/3 \cdot v_{\Phi}/\sqrt{2}$, the top quark mass is set to its SM value, while for the other quarks we chose $\mathcal{O}(1)$~MeV masses. We can see that it is similar to the case where we also set the top quark mass to be negligible, {\it i.e.} $\mathcal{O}(1)$~MeV (``full result w/ $m_t \to 0$", purple, small dashing). The large value of the top quark mass leads to a small increase of the branching fraction for large $m_{Z'}$. Both of these curves show a turnover at small $m_{Z'}$ and recover the Landau-Yang limit for $m_{Z'} \to 0$.

In the curve labelled ``EFT" (blue, medium dashing) we decouple the anomalons ($M \to \infty$) and assume all SM quark masses to vanish ($m_q \to 0$). This becomes a good approximation for small gauge couplings $g_X$, up to a small deviation from the top quark mass visible at higher $m_{Z_B'}$. Note, however, that the EFT curve does not turn over but diverges for $m_{Z_B'} \to 0$. 
Therefore the EFT approximation that decouples the anomalons is valid in an increasing interval, but it breaks down as $m_{Z_B'} \to 0$.

When we neglect the SM fermions entirely, {\it i.e.} not only assume their masses to vanish but also neglect their constant anomaly piece, we have to fix the values of the shift parameters $w$ and $z$. Thus, we take the contribution from the two anomalons that have axial-vector couplings to the $Z_B'$ boson and vector-like couplings to the SM $Z$, and set $w=2z+1$ and $z=-3/2$, which we calculated to be the necessary settings to reproduce the correct Wess-Zumino term. For the branching fraction curve named ``EFT w/o SM fermions" (green, dot-dashed) we furthermore decouple the anomalons, $M \to \infty$. It becomes clear that the contribution from the SM fermions cannot be neglected, since they give large contributions for small as well as large gauge couplings $g_X$ masses and over the whole range of $Z_B'$ masses.

The next simplification we apply is to take the longitudinal polarisation only of the $Z_B'$ (``EFT w/o SM fermions, long. pol.", yellow, large dashing). Still, all SM fermions are neglected, the anomalon masses are sent to infinity, and $w$ and $z$ are set as before. The branching fraction becomes smaller at larger $m_{Z'}$, when comparing to the EFT result without SM fermions including the full polarisation. We expect the GBE to break down here, it is valid at energies $\gg m_{Z'}$. The curve labelled ``GBE" (red, solid) additionally takes the Goldstone mass $m_{Z'}$ to be small compared to the $Z$ boson mass, preventing the branching fraction to fall so quickly as $m_{Z'}$ approaches $m_Z$.

\begin{figure}[htb!]
\begin{center}
\includegraphics[width=\linewidth, angle=0]{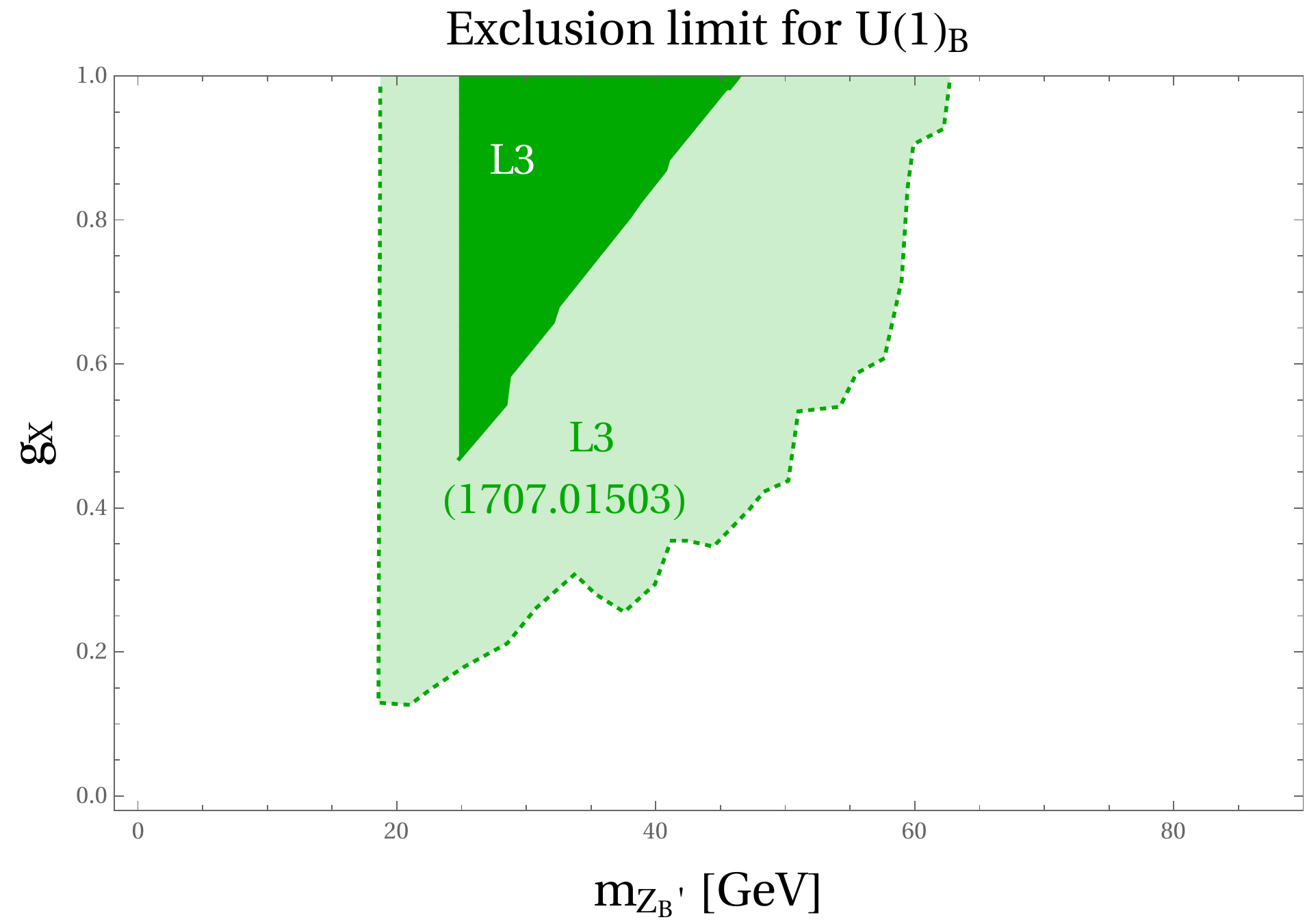} 
\caption{Exclusion limit for $U(1)_B$ in the $g_X$, $m_{Z_B'}$ plane, where the limit resulting from our calculation is shown in solid green, comparing to the limit calculated in Refs.~\cite{Dror:2017ehi, Dror:2017nsg} in transparent green, both using the measurements on the $Z \to (jj)\gamma$ branching fraction from L3~\cite{Adriani:1992zm}. Note that in Refs.~\cite{Dror:2017ehi, Dror:2017nsg} the SM fermions were omitted in the calculation of the width and we furthermore obtain a factor 16 difference to their result when reproducing the GBE ansatz.}
\label{fig:gX_exc_Posp}
\end{center}
\end{figure}

In summary, the most important conclusion is that the constant piece from the SM quarks cannot be neglected.  We have also seen that the EFT approach is generally acceptable for a wide range of $m_{Z_B'}$ masses and small $g_X$ couplings, as long as the SM fermions are included.  Numerically, the large top quark mass actually has a rather small impact. The omission of the constant piece from the SM fermions, however, leads to a break down of the calculation independent of the gauge coupling $g_X$ and the mass of the $Z_B'$.

Taking a Wess-Zumino term as in~\eqnref{GBELang} to represent the anomaly cancelling physics is therefore only valid if the SM fermions are still being considered in the calculations. Then the shift parameters $w$ and $z$ have to be set in the SM fermion induced matrix element to match the choice of the Wilson coefficient $C_B$ and the width has to be calculated adding the contributions of the Wess-Zumino term and the SM fermions coherently.

We remark that if the anomalons were heavy copies of the SM leptons, then their axial-vector coupling lies on the $Z$ vertex instead of the $Z_B'$ vertex and the corresponding $Z \to Z_B' \gamma$ decay width is entirely driven by intragenerational mass splittings.  The corresponding effective operator treatment is then driven by non-decoupling effects of heavy fermions, and since all fermions share the same chiral symmetry breaking scale distinct from $U(1)_B$ breaking, the EFT treatment is valid for generic choices of $U(1)_B$ parameters.

Having analyzed the limitations of the effective Wess-Zumino and furthermore the GBE ansatz, we need to comment on the limit calculation in the $g_X$ versus $m_{Z_B'}$ plane that was obtained in Refs.~\cite{Dror:2017ehi, Dror:2017nsg}, where they use the measurements from the L3 experiment at LEP~\cite{Adriani:1992zm} on the $Z \to (jj)\gamma$ branching fraction. We plot the L3 limit from the full width including anomalons with maximal Yukawa couplings and the correct top quark mass in~\figref{gX_exc} (green solid, ``L3"). Comparing our L3 limit with the one obtained in Refs.~\cite{Dror:2017ehi, Dror:2017nsg} in~\figref{gX_exc_Posp}, we see that for the full calculation the limit becomes significantly weaker.  We note that when we take the width for the GBE we obtained in~\eqnref{GBEwidth}, the limit is too weak to appear on~\figref{gX_exc_Posp}~\footnote{The GBE width we calculate in~\eqnref{GBEwidth} is a factor 16 smaller compared to the result that was stated in Refs.~\cite{Dror:2017ehi, Dror:2017nsg}.}.

\end{appendix}


\bibliographystyle{JHEP}
\bibliography{anomalies}

\providecommand{\href}[2]{#2}\begingroup\raggedright\begin{thebibliography}{10}

\bibitem{Preskill:1990fr}
J.~Preskill, {\it {Gauge anomalies in an effective field theory}},  {\em Annals
  Phys.} {\bf 210} (1991) 323--379.

\bibitem{FileviezPerez:2010gw}
P.~Fileviez~Perez and M.~B. Wise, {\it {Baryon and lepton number as local gauge
  symmetries}},  {\em Phys. Rev. D} {\bf 82} (2010) 011901,
  [\href{http://arxiv.org/abs/1002.1754}{{\tt arXiv:1002.1754}}]. [Erratum:
  Phys.Rev.D 82, 079901 (2010)].

\bibitem{Carena:2004xs}
M.~Carena, A.~Daleo, B.~A. Dobrescu, and T.~M. Tait, {\it {$Z^\prime$ gauge
  bosons at the Tevatron}},  {\em Phys. Rev. D} {\bf 70} (2004) 093009,
  [\href{http://arxiv.org/abs/hep-ph/0408098}{{\tt hep-ph/0408098}}].

\bibitem{Langacker:2008yv}
P.~Langacker, {\it {The Physics of Heavy $Z^\prime$ Gauge Bosons}},  {\em Rev.
  Mod. Phys.} {\bf 81} (2009) 1199--1228,
  [\href{http://arxiv.org/abs/0801.1345}{{\tt arXiv:0801.1345}}].

\bibitem{Dobrescu:2013cmh}
B.~A. Dobrescu and F.~Yu, {\it {Coupling-Mass Mapping of Dijet Peak Searches}},
   {\em Phys. Rev. D} {\bf 88} (2013), no.~3 035021,
  [\href{http://arxiv.org/abs/1306.2629}{{\tt arXiv:1306.2629}}]. [Erratum:
  Phys.Rev.D 90, 079901 (2014)].

\bibitem{Anastasopoulos:2006cz}
P.~Anastasopoulos, M.~Bianchi, E.~Dudas, and E.~Kiritsis, {\it {Anomalies,
  anomalous U(1)'s and generalized Chern-Simons terms}},  {\em JHEP} {\bf 11}
  (2006) 057, [\href{http://arxiv.org/abs/hep-th/0605225}{{\tt
  hep-th/0605225}}].

\bibitem{FileviezPerez:2011pt}
P.~Fileviez~Perez and M.~B. Wise, {\it {Breaking Local Baryon and Lepton Number
  at the TeV Scale}},  {\em JHEP} {\bf 08} (2011) 068,
  [\href{http://arxiv.org/abs/1106.0343}{{\tt arXiv:1106.0343}}].

\bibitem{Cui:2017juz}
Y.~Cui and F.~D'Eramo, {\it {Surprises from complete vector portal theories:
  New insights into the dark sector and its interplay with Higgs physics}},
  {\em Phys. Rev. D} {\bf 96} (2017), no.~9 095006,
  [\href{http://arxiv.org/abs/1705.03897}{{\tt arXiv:1705.03897}}].

\bibitem{Dror:2017ehi}
J.~A. Dror, R.~Lasenby, and M.~Pospelov, {\it {New constraints on light vectors
  coupled to anomalous currents}},  {\em Phys. Rev. Lett.} {\bf 119} (2017),
  no.~14 141803, [\href{http://arxiv.org/abs/1705.06726}{{\tt
  arXiv:1705.06726}}].

\bibitem{Arcadi:2017jqd}
G.~Arcadi, P.~Ghosh, Y.~Mambrini, M.~Pierre, and F.~S. Queiroz, {\it {$Z'$
  portal to Chern-Simons Dark Matter}},  {\em JCAP} {\bf 11} (2017) 020,
  [\href{http://arxiv.org/abs/1706.04198}{{\tt arXiv:1706.04198}}].

\bibitem{Dror:2017nsg}
J.~A. Dror, R.~Lasenby, and M.~Pospelov, {\it {Dark forces coupled to
  nonconserved currents}},  {\em Phys. Rev.} {\bf D96} (2017), no.~7 075036,
  [\href{http://arxiv.org/abs/1707.01503}{{\tt arXiv:1707.01503}}].

\bibitem{Ismail:2017ulg}
A.~Ismail, A.~Katz, and D.~Racco, {\it {On dark matter interactions with the
  Standard Model through an anomalous $Z'$}},  {\em JHEP} {\bf 10} (2017) 165,
  [\href{http://arxiv.org/abs/1707.00709}{{\tt arXiv:1707.00709}}].

\bibitem{Ismail:2017fgq}
A.~Ismail and A.~Katz, {\it {Anomalous $Z'$ and diboson resonances at the
  LHC}},  {\em JHEP} {\bf 04} (2018) 122,
  [\href{http://arxiv.org/abs/1712.01840}{{\tt arXiv:1712.01840}}].

\bibitem{Dror:2018wfl}
J.~A. Dror, R.~Lasenby, and M.~Pospelov, {\it {Light vectors coupled to bosonic
  currents}},  {\em Phys. Rev. D} {\bf 99} (2019), no.~5 055016,
  [\href{http://arxiv.org/abs/1811.00595}{{\tt arXiv:1811.00595}}].

\bibitem{Batra:2005rh}
P.~Batra, B.~A. Dobrescu, and D.~Spivak, {\it {Anomaly-free sets of fermions}},
   {\em J. Math. Phys.} {\bf 47} (2006) 082301,
  [\href{http://arxiv.org/abs/hep-ph/0510181}{{\tt hep-ph/0510181}}].

\bibitem{Allanach:2018vjg}
B.~Allanach, J.~Davighi, and S.~Melville, {\it {An Anomaly-free Atlas: charting
  the space of flavour-dependent gauged $U(1)$ extensions of the Standard
  Model}},  {\em JHEP} {\bf 02} (2019) 082,
  [\href{http://arxiv.org/abs/1812.04602}{{\tt arXiv:1812.04602}}]. [Erratum:
  JHEP 08, 064 (2019)].

\bibitem{Costa:2019zzy}
D.~B. Costa, B.~A. Dobrescu, and P.~J. Fox, {\it {General Solution to the U(1)
  Anomaly Equations}},  {\em Phys. Rev. Lett.} {\bf 123} (2019), no.~15 151601,
  [\href{http://arxiv.org/abs/1905.13729}{{\tt arXiv:1905.13729}}].

\bibitem{Costa:2020dph}
D.~B. Costa, B.~A. Dobrescu, and P.~J. Fox, {\it {Chiral Abelian gauge theories
  with few fermions}},  {\em Phys. Rev. D} {\bf 101} (2020) 095032,
  [\href{http://arxiv.org/abs/2001.11991}{{\tt arXiv:2001.11991}}].

\bibitem{Kribs:2007nz}
G.~D. Kribs, T.~Plehn, M.~Spannowsky, and T.~M.~P. Tait, {\it {Four generations
  and Higgs physics}},  {\em Phys. Rev.} {\bf D76} (2007) 075016,
  [\href{http://arxiv.org/abs/0706.3718}{{\tt arXiv:0706.3718}}].

\bibitem{Khachatryan:2016vau}
{\bf ATLAS, CMS} Collaboration, G.~Aad et~al., {\it {Measurements of the Higgs
  boson production and decay rates and constraints on its couplings from a
  combined ATLAS and CMS analysis of the LHC pp collision data at $ \sqrt{s}=7
  $ and 8 TeV}},  {\em JHEP} {\bf 08} (2016) 045,
  [\href{http://arxiv.org/abs/1606.02266}{{\tt arXiv:1606.02266}}].

\bibitem{Wess:1971yu}
J.~Wess and B.~Zumino, {\it {Consequences of anomalous Ward identities}},  {\em
  Phys. Lett. B} {\bf 37} (1971) 95--97.

\bibitem{Adler:1969gk}
S.~L. Adler, {\it {Axial vector vertex in spinor electrodynamics}},  {\em Phys.
  Rev.} {\bf 177} (1969) 2426--2438. [,241(1969)].

\bibitem{Bell:1969ts}
J.~S. Bell and R.~Jackiw, {\it {A PCAC puzzle: $\pi^0 \to \gamma \gamma$ in the
  $\sigma$ model}},  {\em Nuovo Cim.} {\bf A60} (1969) 47--61.

\bibitem{Dedes:2012me}
A.~Dedes and K.~Suxho, {\it {Heavy Fermion Non-Decoupling Effects in Triple
  Gauge Boson Vertices}},  {\em Phys. Rev.} {\bf D85} (2012) 095024,
  [\href{http://arxiv.org/abs/1202.4940}{{\tt arXiv:1202.4940}}].

\bibitem{DHoker:1984mif}
E.~D'Hoker and E.~Farhi, {\it {Decoupling a Fermion in the Standard Electroweak
  Theory}},  {\em Nucl. Phys. B} {\bf 248} (1984) 77.

\bibitem{DHoker:1984izu}
E.~D'Hoker and E.~Farhi, {\it {Decoupling a Fermion Whose Mass Is Generated by
  a Yukawa Coupling: The General Case}},  {\em Nucl. Phys. B} {\bf 248} (1984)
  59--76.

\bibitem{Shifman:1979eb}
M.~A. Shifman, A.~Vainshtein, M.~Voloshin, and V.~I. Zakharov, {\it {Low-Energy
  Theorems for Higgs Boson Couplings to Photons}},  {\em Sov. J. Nucl. Phys.}
  {\bf 30} (1979) 711--716.

\bibitem{Glashow:1970gm}
S.~Glashow, J.~Iliopoulos, and L.~Maiani, {\it {Weak Interactions with
  Lepton-Hadron Symmetry}},  {\em Phys. Rev. D} {\bf 2} (1970) 1285--1292.

\bibitem{Dobrescu:2014fca}
B.~A. Dobrescu and C.~Frugiuele, {\it {Hidden GeV-scale interactions of
  quarks}},  {\em Phys. Rev. Lett.} {\bf 113} (2014) 061801,
  [\href{http://arxiv.org/abs/1404.3947}{{\tt arXiv:1404.3947}}].

\bibitem{Appelquist:1987cf}
T.~Appelquist and M.~S. Chanowitz, {\it {Unitarity Bound on the Scale of
  Fermion Mass Generation}},  {\em Phys. Rev. Lett.} {\bf 59} (1987) 2405.
  [Erratum: Phys. Rev. Lett.60,1589(1988)].

\bibitem{Weinberg:1996kr}
S.~Weinberg, {\em {The quantum theory of fields. Vol. 2: Modern applications}}.
\newblock Cambridge University Press, 2013.

\bibitem{Rosenberg:1962pp}
L.~Rosenberg, {\it {Electromagnetic interactions of neutrinos}},  {\em Phys.
  Rev.} {\bf 129} (1963) 2786--2788.

\bibitem{Treiman:1986ep}
S.~Treiman, E.~Witten, R.~Jackiw, and B.~Zumino, {\em {CURRENT ALGEBRA AND
  ANOMALIES}}.
\newblock 9, 1986.

\bibitem{Mathematica}
W.~R. Inc., ``Mathematica, {V}ersion 12.1.''
\newblock Champaign, IL, 2020.

\bibitem{Patel:2015tea}
H.~H. Patel, {\it {Package-X: A Mathematica package for the analytic
  calculation of one-loop integrals}},  {\em Comput. Phys. Commun.} {\bf 197}
  (2015) 276--290, [\href{http://arxiv.org/abs/1503.01469}{{\tt
  arXiv:1503.01469}}].

\bibitem{Patel:2016fam}
H.~H. Patel, {\it {Package-X 2.0: A Mathematica package for the analytic
  calculation of one-loop integrals}},  {\em Comput. Phys. Commun.} {\bf 218}
  (2017) 66--70, [\href{http://arxiv.org/abs/1612.00009}{{\tt
  arXiv:1612.00009}}].

\bibitem{Passarino:1978jh}
G.~Passarino and M.~J.~G. Veltman, {\it {One Loop Corrections for e+ e-
  Annihilation Into mu+ mu- in the Weinberg Model}},  {\em Nucl. Phys.} {\bf
  B160} (1979) 151--207.

\bibitem{Landau:1948kw}
L.~D. Landau, {\it {On the angular momentum of a system of two photons}},  {\em
  Dokl. Akad. Nauk SSSR} {\bf 60} (1948), no.~2 207--209.

\bibitem{Yang:1950rg}
C.-N. Yang, {\it {Selection Rules for the Dematerialization of a Particle Into
  Two Photons}},  {\em Phys. Rev.} {\bf 77} (1950) 242--245.

\bibitem{Adeva:1991dw}
{\bf L3} Collaboration, B.~Adeva et~al., {\it {Search for narrow high mass
  resonances in radiative decays of the Z0}},  {\em Phys. Lett. B} {\bf 262}
  (1991) 155--162.

\bibitem{Adriani:1992zm}
{\bf L3} Collaboration, O.~Adriani et~al., {\it {Isolated hard photon emission
  in hadronic Z0 decays}},  {\em Phys. Lett. B} {\bf 292} (1992) 472--484.

\bibitem{Tulin:2014tya}
S.~Tulin, {\it {New weakly-coupled forces hidden in low-energy QCD}},  {\em
  Phys. Rev. D} {\bf 89} (2014), no.~11 114008,
  [\href{http://arxiv.org/abs/1404.4370}{{\tt arXiv:1404.4370}}].

\bibitem{Achard:2001qw}
{\bf L3} Collaboration, P.~Achard et~al., {\it {Search for heavy neutral and
  charged leptons in $e^{+} e^{-}$ annihilation at LEP}},  {\em Phys. Lett.}
  {\bf B517} (2001) 75--85, [\href{http://arxiv.org/abs/hep-ex/0107015}{{\tt
  hep-ex/0107015}}].

\bibitem{Heister:2002mn}
{\bf ALEPH} Collaboration, A.~Heister et~al., {\it {Search for charginos nearly
  mass degenerate with the lightest neutralino in e+ e- collisions at
  center-of-mass energies up to 209-GeV}},  {\em Phys. Lett. B} {\bf 533}
  (2002) 223--236, [\href{http://arxiv.org/abs/hep-ex/0203020}{{\tt
  hep-ex/0203020}}].

\bibitem{Egana-Ugrinovic:2018roi}
D.~Egana-Ugrinovic, M.~Low, and J.~T. Ruderman, {\it {Charged Fermions Below
  100 GeV}},  {\em JHEP} {\bf 05} (2018) 012,
  [\href{http://arxiv.org/abs/1801.05432}{{\tt arXiv:1801.05432}}].

\bibitem{Acton:1991dq}
{\bf OPAL} Collaboration, P.~Acton et~al., {\it {A Measurement of photon
  radiation in lepton pair events from Z0 decays}},  {\em Phys. Lett. B} {\bf
  273} (1991) 338--354.

\bibitem{Cacciari:2015ela}
M.~Cacciari, L.~Del~Debbio, J.~R. Espinosa, A.~D. Polosa, and M.~Testa, {\it {A
  note on the fate of the Landau--Yang theorem in non-Abelian gauge theories}},
   {\em Phys. Lett. B} {\bf 753} (2016) 476--481,
  [\href{http://arxiv.org/abs/1509.07853}{{\tt arXiv:1509.07853}}].

\bibitem{Djouadi:2005gi}
A.~Djouadi, {\it {The Anatomy of electro-weak symmetry breaking. I: The Higgs
  boson in the standard model}},  {\em Phys. Rept.} {\bf 457} (2008) 1--216,
  [\href{http://arxiv.org/abs/hep-ph/0503172}{{\tt hep-ph/0503172}}].

\bibitem{Aaboud:2018xdt}
{\bf ATLAS} Collaboration, M.~Aaboud et~al., {\it {Measurements of Higgs boson
  properties in the diphoton decay channel with 36 fb$^{-1}$ of $pp$ collision
  data at $\sqrt{s} = 13$ TeV with the ATLAS detector}},  {\em Phys. Rev.} {\bf
  D98} (2018) 052005, [\href{http://arxiv.org/abs/1802.04146}{{\tt
  arXiv:1802.04146}}].

\bibitem{Schwaller:2013baa}
P.~Schwaller and J.~Zurita, {\it {Compressed electroweakino spectra at the
  LHC}},  {\em JHEP} {\bf 03} (2014) 060,
  [\href{http://arxiv.org/abs/1312.7350}{{\tt arXiv:1312.7350}}].

\bibitem{Aad:2019vnb}
{\bf ATLAS} Collaboration, G.~Aad et~al., {\it {Search for electroweak
  production of charginos and sleptons decaying into final states with two
  leptons and missing transverse momentum in $\sqrt{s}=13$ TeV $pp$ collisions
  using the ATLAS detector}},  {\em Eur. Phys. J. C} {\bf 80} (2020), no.~2
  123, [\href{http://arxiv.org/abs/1908.08215}{{\tt arXiv:1908.08215}}].

\bibitem{Aad:2019qnd}
{\bf ATLAS} Collaboration, G.~Aad et~al., {\it {Searches for electroweak
  production of supersymmetric particles with compressed mass spectra in
  $\sqrt{s}=$ 13 TeV $pp$ collisions with the ATLAS detector}},  {\em Phys.
  Rev. D} {\bf 101} (2020), no.~5 052005,
  [\href{http://arxiv.org/abs/1911.12606}{{\tt arXiv:1911.12606}}].

\bibitem{Sirunyan:2019sgo}
{\bf CMS} Collaboration, A.~M. Sirunyan et~al., {\it {Search for Low-Mass
  Quark-Antiquark Resonances Produced in Association with a Photon at $\sqrt
  {s}$ =13 TeV}},  {\em Phys. Rev. Lett.} {\bf 123} (2019), no.~23 231803,
  [\href{http://arxiv.org/abs/1905.10331}{{\tt arXiv:1905.10331}}].

\bibitem{Sirunyan:2019vxa}
{\bf CMS} Collaboration, A.~M. Sirunyan et~al., {\it {Search for low mass
  vector resonances decaying into quark-antiquark pairs in proton-proton
  collisions at $\sqrt{s}=$ 13 TeV}},  {\em Phys. Rev. D} {\bf 100} (2019),
  no.~11 112007, [\href{http://arxiv.org/abs/1909.04114}{{\tt
  arXiv:1909.04114}}].

\bibitem{Albrecht:1986ec}
{\bf ARGUS} Collaboration, H.~Albrecht et~al., {\it {An Upper Limit for Two Jet
  Production in Direct $\Upsilon$ (1s) Decays}},  {\em Z. Phys. C} {\bf 31}
  (1986) 181.

\bibitem{Aranda:1998fr}
A.~Aranda and C.~D. Carone, {\it {Limits on a light leptophobic gauge boson}},
  {\em Phys. Lett. B} {\bf 443} (1998) 352--358,
  [\href{http://arxiv.org/abs/hep-ph/9809522}{{\tt hep-ph/9809522}}].

\bibitem{Carone:1994aa}
C.~D. Carone and H.~Murayama, {\it {Possible light U(1) gauge boson coupled to
  baryon number}},  {\em Phys. Rev. Lett.} {\bf 74} (1995) 3122--3125,
  [\href{http://arxiv.org/abs/hep-ph/9411256}{{\tt hep-ph/9411256}}].

\bibitem{Bailey:1994qv}
D.~C. Bailey and S.~Davidson, {\it {Is there a vector boson coupling to baryon
  number?}},  {\em Phys. Lett. B} {\bf 348} (1995) 185--189,
  [\href{http://arxiv.org/abs/hep-ph/9411355}{{\tt hep-ph/9411355}}].

\bibitem{Carone:1995pu}
C.~D. Carone and H.~Murayama, {\it {Realistic models with a light U(1) gauge
  boson coupled to baryon number}},  {\em Phys. Rev. D} {\bf 52} (1995)
  484--493, [\href{http://arxiv.org/abs/hep-ph/9501220}{{\tt hep-ph/9501220}}].

\bibitem{Blondel:2018mad}
A.~Blondel et~al., {\it {Standard model theory for the FCC-ee Tera-Z stage}},
  in {\em {Mini Workshop on Precision EW and QCD Calculations for the FCC
  Studies : Methods and Techniques}}, vol.~3/2019 of {\em CERN Yellow Reports:
  Monographs}, (Geneva), CERN, 9, 2018.
\newblock \href{http://arxiv.org/abs/1809.01830}{{\tt arXiv:1809.01830}}.

\bibitem{ALEPH:2005ab}
{\bf ALEPH, DELPHI, L3, OPAL, SLD, LEP Electroweak Working Group, SLD
  Electroweak Group, SLD Heavy Flavour Group} Collaboration, S.~Schael et~al.,
  {\it {Precision electroweak measurements on the $Z$ resonance}},  {\em Phys.
  Rept.} {\bf 427} (2006) 257--454,
  [\href{http://arxiv.org/abs/hep-ex/0509008}{{\tt hep-ex/0509008}}].

\bibitem{Binosi:2008ig}
D.~Binosi, J.~Collins, C.~Kaufhold, and L.~Theussl, {\it {JaxoDraw: A Graphical
  user interface for drawing Feynman diagrams. Version 2.0 release notes}},
  {\em Comput. Phys. Commun.} {\bf 180} (2009) 1709--1715,
  [\href{http://arxiv.org/abs/0811.4113}{{\tt arXiv:0811.4113}}].

\end{thebibliography}\endgroup

\end{document}